\documentclass[prb, twocolumn, nofootinbib]{revtex4-2}

\usepackage{bibunits}
\defaultbibliographystyle{apsrev4-2}
\renewcommand{\citenum}[1]{\onlinecite{#1}}

\usepackage{amssymb}
\usepackage{amsmath}
\usepackage{bbold}
\usepackage{mathrsfs} 
\usepackage{graphicx}
\usepackage{bm}
\usepackage{tabularx}
\usepackage[x11names]{xcolor}
\usepackage[unicode=true,colorlinks=true]{hyperref} 
\hypersetup{urlcolor = blue, linkcolor=blue }

\newcommand{\ket}[1]{\left\vert #1 \right\rangle}

\newcommand{\braket}[3]{\left\langle #1 \left| #2 \right| #3 \right\rangle}

\begin{document}
\title{Giant optical orientation of exciton spins in lead halide perovskite crystals}

\author{Natalia~E.~Kopteva$^{1}$, Dmitri~R.~Yakovlev$^{1}$, Ey\"up~Yalcin$^{1}$, Ilya~A.~Akimov$^{1}$,  Mikhail~O.~Nestoklon$^{1}$, Mikhail~M.~Glazov$^2$, Mladen~Kotur$^{1}$, Dennis~Kudlacik$^{1}$, Evgeny~A.~Zhukov$^{1}$, Erik~Kirstein$^{1}$, Oleh~Hordiichuk$^{3,4}$, Dmitry~N.~Dirin$^{3,4}$, Maksym~V.~Kovalenko$^{3,4}$, and Manfred~Bayer$^{1}$}

\affiliation{$^{1}$Experimentelle Physik 2, Technische Universit\"at Dortmund, 44227 Dortmund, Germany}
\affiliation{$^{2}$Ioffe Institute, Russian Academy of Sciences, 194021 St. Petersburg, Russia}
\affiliation{$^{3}$Laboratory of Inorganic Chemistry, Department of Chemistry and Applied Biosciences,  ETH Z\"{u}rich, CH-8093 Z\"{u}rich, Switzerland}
\affiliation{$^{4}$Laboratory for Thin Films and Photovoltaics, Empa-Swiss Federal Laboratories for Materials Science and Technology, CH-8600 D\"{u}bendorf, Switzerland }

\date{\today}
\makeatletter
\newenvironment{mywidetext}{%
  \par\ignorespaces
  \setbox\widetext@top\vbox{%
   \hb@xt@\hsize{%
    \leaders\hrule\hfil
    \vrule\@height6\p@
   }%
  }%
  \setbox\widetext@bot\hb@xt@\hsize{%
    \vrule\@depth6\p@
    \leaders\hrule\hfil
  }%
  \onecolumngrid
  \vskip10\p@
  \dimen@\ht\widetext@top\advance\dimen@\dp\widetext@top
  \cleaders\box\widetext@top\vskip\dimen@
  \vskip6\p@
  \prep@math@patch
}{%
  \par
  \vskip6\p@
  \setbox\widetext@bot\vbox{%
   \hb@xt@\hsize{\hfil\box\widetext@bot}%
  }%
  \dimen@\ht\widetext@bot\advance\dimen@\dp\widetext@bot
  \vskip\dimen@
  \vskip8.5\p@
  \twocolumngrid\global\@ignoretrue
  \@endpetrue
}%
\makeatother

\begin{abstract}
Optical orientation of carrier spins by circularly polarized light is the basis of spin physics in semiconductors. Here, we demonstrate strong optical orientation of 85\%, approaching the ultimate limit of unity, for excitons in FA$_{0.9}$Cs$_{0.1}$PbI$_{2.8}$Br$_{0.2}$ lead halide perovskite bulk crystals. Time-resolved photoluminescence allows us to distinguish excitons with 60~ps lifetime from electron-hole recombination in the spin dynamics detected via coherent spin quantum beats in magnetic field. We reveal electron-hole spin correlations through linear polarization beats after circularly polarized excitation. Detuning of the excitation energy from the exciton resonance up to 0.5~eV does not reduce the optical orientation, evidencing clean chiral selection rules in agreement with atomistic calculations, and suppressed spin relaxation of electrons and holes even with large kinetic energies.   \\
\end{abstract}

\maketitle

\begin{bibunit}

Lead halide perovskite semiconductors are known for their exceptional photovoltaic efficiency~\cite{jena2019,nrel2021} and optoelectronic properties~\cite{Vinattieri2021_book,Vardeny2022_book}. Their simple fabrication technology makes them attractive for applications as solar cells or light emitting diodes. They demonstrate also remarkable spin features facilitating spintronic applications~\cite{Vardeny2022_book,wang2019,ning2020,kim2021}. The spin physics of halide perovskite semiconductors still is an emerging research field, which exploits experimental techniques and physical concepts developed for spins in conventional semiconductors~\cite{Spin_book_2017}. Most of spin-dependent optical techniques work well for perovskite crystals, polycrystalline films, nanocrystals, and two-dimensional materials. These are: optical orientation~\cite{Giovanni2015,Nestoklon-PRB2018}, optical alignment~\cite{Nestoklon-PRB2018}, polarized emission in magnetic field~\cite{zhang2015,zhang_field-induced_2018}, time-resolved Faraday/Kerr rotation~\cite{odenthal2017,belykh2019}, spin-flip Raman scattering~\cite{kirstein2021nc,Harkort_2D_2023}, and optically-detected nuclear magnetic resonance~\cite{kirstein2021}. Universal dependences of the electron, hole, and exciton Land\`e $g$-factors on the band gap energy have been established~\cite{kirstein2021nc,Kopteva_gX_2023}. The reported spin dynamics cover huge time ranges from a few picoseconds at room temperature~\cite{Giovanni2015,strohmair2020} up to tens nanoseconds for the spin coherence~\cite{Kirstein_SML_2023} and spin dephasing~\cite{kirstein2021} times and further up to sub-milliseconds for the longitudinal spin relaxation times~\cite{Belykh2022} at cryogenic temperatures.

Optical orientation is a fundamental phenomenon in spin physics~\cite{OO_book, Spin_book_2017}, where circularly polarized photons generate spin oriented  excitons and charge carriers, whose spin polarization can be monitored, also dynamically, via polarized photoluminescence, Faraday/Kerr rotation, spin-dependent photocurrents, etc. Optical pulses with subpicosecond duration can be used for ultrafast spin orientation, manipulation, and read out - operations generally required also for quantum information technologies. For lead halide perovskites, optical orientation under pulsed excitation was used to trigger spin dynamics.\cite{Giovanni2015,strohmair2020,zhou2020,odenthal2017} All-optical manipulation of carrier spins in singly-charged CsPbBr$_3$ nanocrystals recently was demonstrated at room temperature.\cite{Lin2022} Using continuous-wave excitation, only small degrees of optical orientation measured via circular polarization of photoluminescence were so far reported for MAPbBr$_3$ polycrystalline films (degree of 3.1\%~\cite{wang2019} at 10~K temperature, 2\%~\cite{Wang2018} and 8\%~\cite{wu2019} at 77~K), for MAPbI$_3$ (0.15\%~\cite{Wang2018} at 77~K), and for  CsPbI$_3$ NCs (4\%~\cite{Nestoklon-PRB2018} at 2~K). On the other hand, the electronic band structure and the resulting selection rules for optical transitions should allow 100\% of carrier spin polarization reflected by 100\% polarized luminescence, compared with the limit of 25\% optical orientation in conventional III-V and II-VI semiconductors. 
Thus, it has remained a challenge to examine the maximum achievable optical orientation in perovskites and to identify the limiting mechanisms involved in spin generation and spin relaxation.

\begin{figure*}[hbt]
\begin{center}
\includegraphics[width = \linewidth]{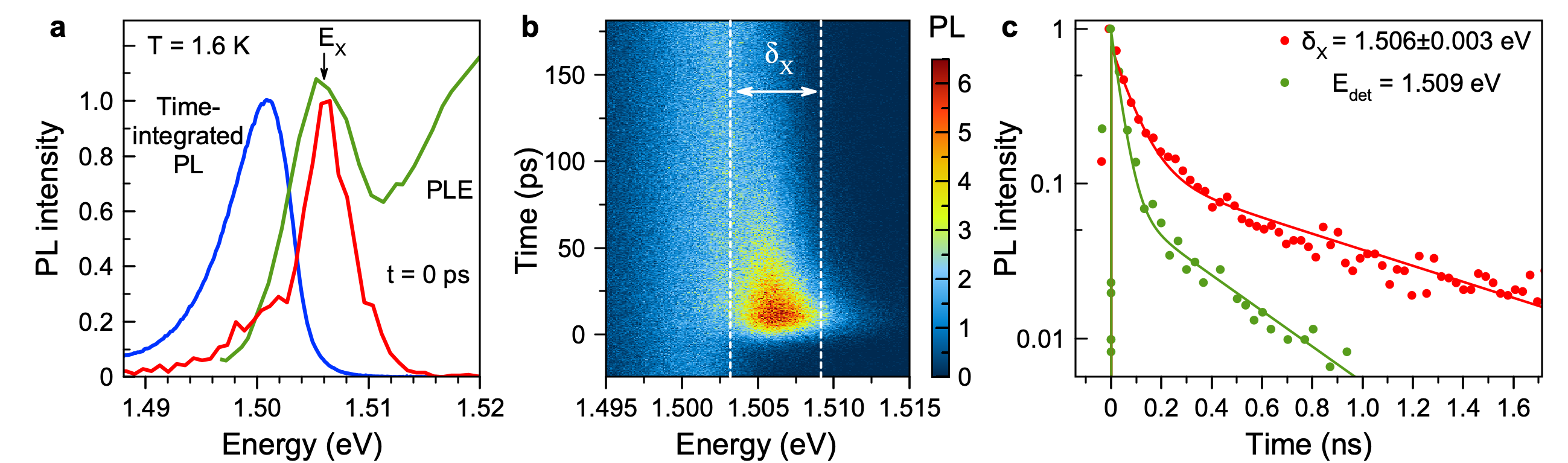}
\caption{\label{fig1} Exciton in bulk FA$_{0.9}$Cs$_{0.1}$PbI$_{2.8}$Br$_{0.2}$ crystal. 
(a) Time-integrated photoluminescence spectrum (blue line) excited at $E_\text{exc} = 1.669$~eV photon energy, using $P = 10$~mW/cm$^2$ laser power. $T = 1.6$~K. Photoluminescence excitation spectrum (green line) detected at $E_\text{det} = 1.496$~eV. $E_\text{X}$ denotes the exciton resonance. PL spectrum at the moment of excitation $t = 0$\,ps for pulsed excitation (red line). (b) Contour plot of time-resolved photoluminescence excited with 200\,fs laser pulses. (c) Recombination dynamics detected at $E_\text{det} = 1.509$\,eV (green) and $1.506$\,eV integrated over the $\delta_\text{X}$ spectral range (red). Lines show bi-exponential fits with decay times: $\tau_{\rm R1} = 55$\,ps and $\tau_{\rm R2} = 840$\,ps for $E_\text{det} = 1.506$\,eV, and $\tau_{\rm R1} = 35$\,ps and $\tau_{\rm R2} = 380$\,ps for $E_\text{det} = 1.509$\,eV. }
\end{center}
\end{figure*}

In this paper we demonstrate that a very high degree of optical orientation up to 85\% can be achieved for excitons in FA$_{0.9}$Cs$_{0.1}$PbI$_{2.8}$Br$_{0.2}$ perovskite crystals. It is surprisingly robust against detuning of the laser excitation energy from the exciton resonance by up to 0.5~eV, evidencing the suppression of the carrier spin relaxation mechanisms typical for conventional III-V and II-VI semiconductors. Atomistic calculations based on density functional theory and empirical tight-binding model calculations accounting for the spin-dependent optical matrix elements at large carrier wave vectors support these observations. Time-resolved photoluminescence allows us to distinguish the excitons with 60~ps lifetime from electron-hole recombination in the spin dynamics detected as coherent spin beats in magnetic field, induced by the circularly polarized excitation and detected in linear or circular polarization, respectively. In the former case, the importance of electron-hole spin correlations is revealed.

\section*{Results}

We choose for this study a bulk single crystal of the FA$_{0.9}$Cs$_{0.1}$PbI$_{2.8}$Br$_{0.2}$ hybrid organic-inorganic lead halide perovskite with high structural quality and small inhomogeneous broadening of the exciton resonance. The small additions of Cs and Br to the basic FAPbI$_3$ composition allows one to set the tolerance factor close to 1. Therefore, this crystal keeps cubic lattice symmetry also at cryogenic temperatures, as confirmed by isotropic electron and hole $g$-factors measured at $T=1.6$~K.\cite{kirstein2021nc}

The optical properties of the studied crystal are illustrated in Fig.~\ref{fig1}a. At the temperature of $T = 1.6$~K the exciton resonance is seen at 1.506~eV in the photoluminescence excitation (PLE) spectrum. The exciton binding energy should be close to the 14~meV for FAPbI$_3$,\cite{galkowski2016} which gives us the band gap energy of $E_g=1.520$~eV in the studied crystal. The time-integrated photoluminescence (PL) spectrum measured under pulsed excitation shows a line with the maximum at 1.501~eV and the full width at half maximum of 5~meV. The recombination dynamics of this line cover a large temporal range from 700~ps to 44~$\mu$s with a large spectral dispersion (SI, S2), indicating a multitude of recombination processes including that of spatially separated charge carriers~\cite{kirstein2021}. The coherent spin dynamics of resident electrons and holes following their optical orientation in such crystals show nanosecond spin dephasing times~\cite{kirstein2021}.

Here, we focus on the spin properties of excitons with short recombination times, for which we use time-resolved photoluminescence (TRPL) recorded with a streak camera to isolate the exciton signals. The PL dynamics are shown as a color map in Fig.~\ref{fig1}b. Right after the photogeneration at $t = 0$\,ps, the emission has its spectral maximum at 1.506~eV, equal to the exciton resonance in the PLE spectrum at $E_\text{X} = 1.506$~eV (red line in Fig.~\ref{fig1}a). The spectrally-integrated exciton emission (red line in Figure~\ref{fig1}c) is observable on time scales up to 1.7~ns, showing a double exponential decay. The fast decay time $\tau_\text{R1} = 55$\,ps is assigned to exciton recombination, and the longer one $\tau_\text{R2} = 840$~\,ps to recombination of spatially separated electrons and holes. It is a specific of the lead halide perovskites that these processes overlap spectrally, which complicates the interpretation of the recombination and spin dynamics.\cite{belykh2019,kirstein2021} The dependences of the exciton and electron-hole pair recombination times on temperature and excitation power are given in SI, S4.

\begin{figure*}[hbt]
\begin{center}
\includegraphics[width =\linewidth]{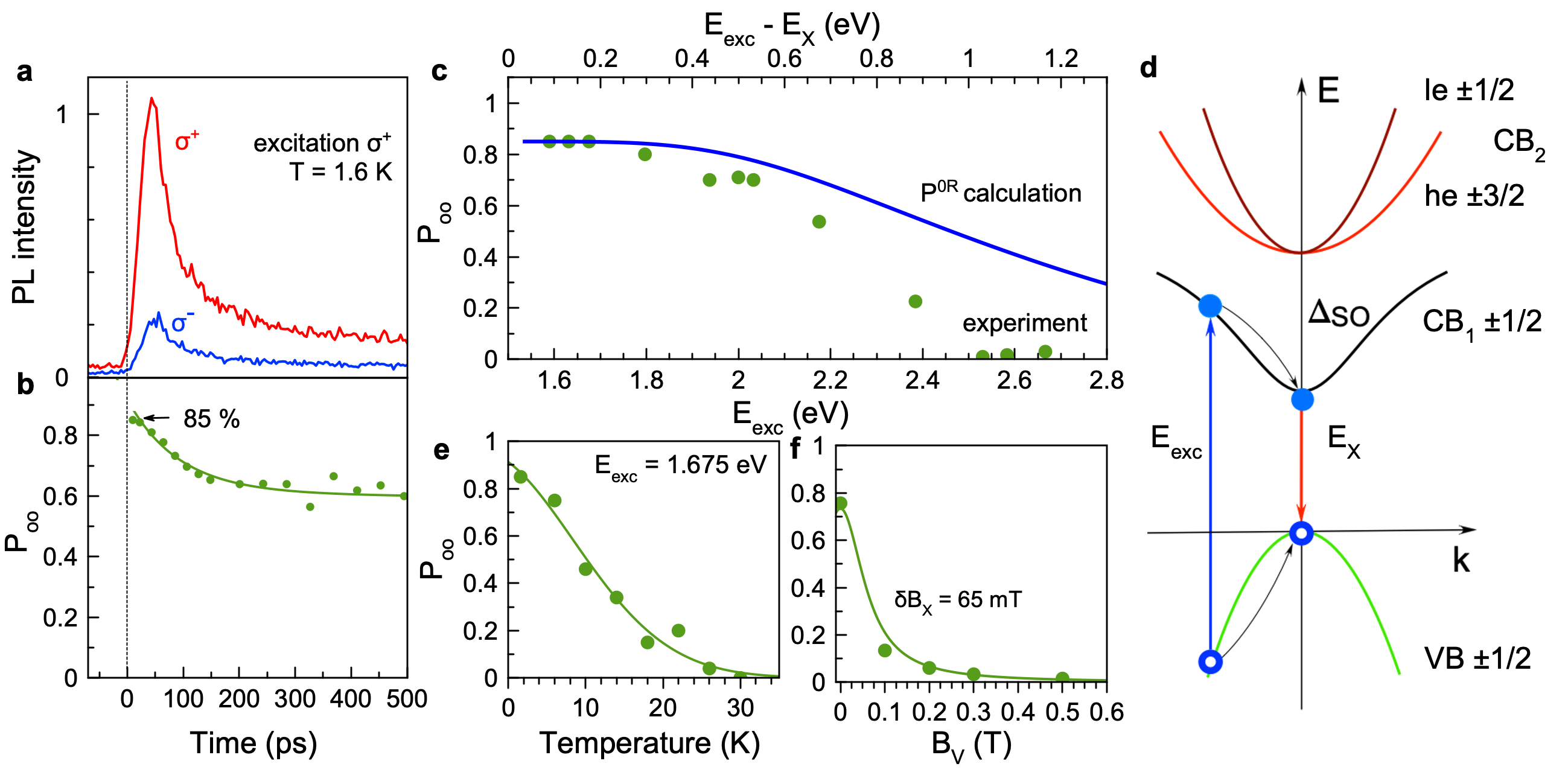}
\caption{\label{fig3} Optical orientation of exciton spins detected at the exciton energy of 1.506~eV.
(a) PL dynamics detected in $\sigma^+$ (red line) and $\sigma^-$ (blue line) polarization for $\sigma^+$ excitation at $E_\text{exc} = 1.669$~eV and $P = 10$~mW/cm$^2$. $T = 1.6$~K. (b) Dynamics of optical orientation degree $P_{\rm oo}(t)$. Line is an exponential fit between the 0.85 and 0.60 levels with the decay time of 100~ps. (c) Dependence of  $P_{\rm oo}(t=0)$ on excitation energy $E_\text{exc}$ (symbols).  Upper axis shows the detuning from the exciton resonance $E_\text{exc} - E_\text{X}$. Blue line is theoretical curve $P^{\rm 0R}$ from Fig.~\ref{fig5}b multiplied by the depolarization
factor 0.85 to match the experimental value of $P_{\rm oo} = 0.85$ at small detunings, see SI, Fig.~S6b. (d) Sketch of the band structure of lead halide perovskites with cubic symmetry. VB and CB$_1$ denote the valence and conduction bands with electron and hole spins $\pm 1/2$. The CB$_2$ band consisting of the heavy (he) and light (le) electron subbands is shifted from CB$_1$ by the spin-orbit splitting $\Delta_{SO}$.  (e) Temperature dependence of $P_{\rm oo}(t=0)$ (symbols). Line is guide to the eye.  (f) $P_\text{oo}$ dependence on the magnetic field applied in the Voigt geometry ($B_\text{V}$) for $\sigma^+$ excitation at $E_\text{exc} = 1.675$\,eV and  $P = 30$\,mW/cm$^2$. $T = 1.6$~K. Each point is obtained by integration of the PL dynamics over 2~ns. Line is a fit with Eq.~(S16), assuming $\delta B_\text{X} = 65$~mT.}
\end{center}
\end{figure*}

\subsection*{Optical orientation of exciton spins}

\begin{figure*}[htb]
\begin{center}
\includegraphics[width = 16cm]{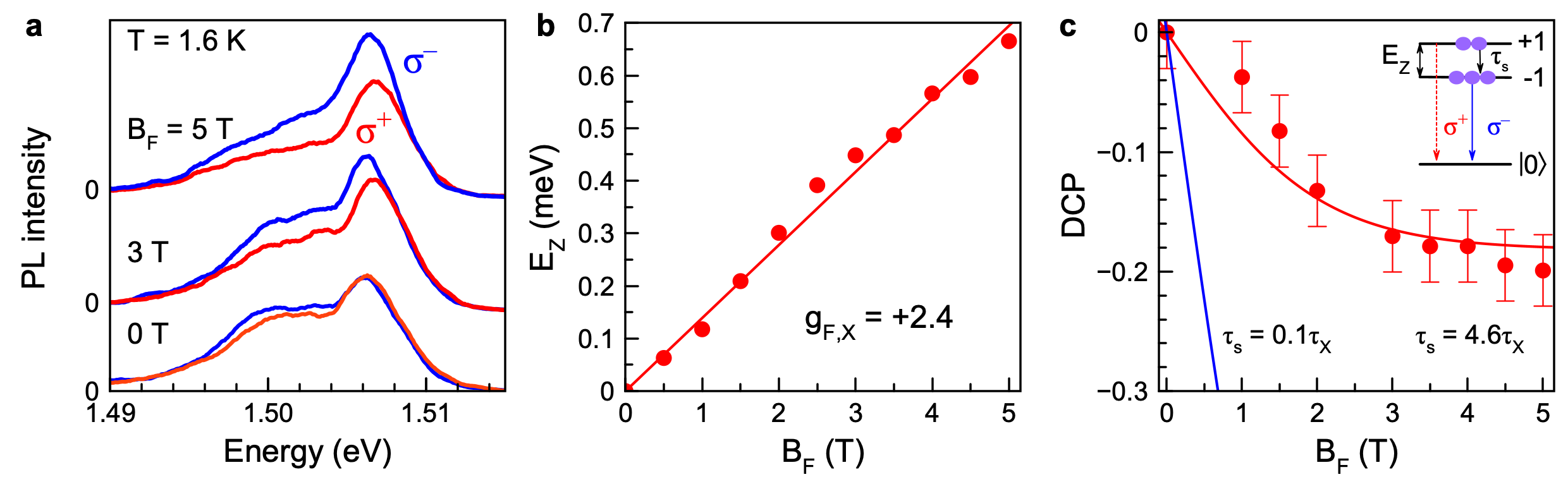}
\caption{\label{fig2} 
Exciton Zeeman splitting and polarization in Faraday magnetic field at $T = 1.6$\,K. (a) PL spectra integrated over $\tau_\text{X}$ in $\sigma^+$ (red) and $\sigma^-$ (blue) polarization in the longitudinal magnetic field $B_\text{F}=5$T. The exciting laser is linearly polarized,  $E_\text{exc} = 1.669$\,eV and $P = 10$\,mW/cm$^2$. (b) Exciton Zeeman splitting as a function of $B_\text{F}$ (symbols). Line is a $B_\text{F}$-linear fit with $g_\text{F,X} = +2.4$. (c) Degree of circular polarization dependence on $B_\text{F}$ (red circles). Red line is a fit using Eq.~\eqref{eq2} with $T = 1.6$\,K and $\tau_{\text{s}} = 4.6\tau_{\text{X}}$. Blue line is the calculated result for the condition $\tau_{\text{s}} = 0.1\tau_{\text{X}}$ at $T = 1.6$\,K. Inset shows schematically the populations of the excitons on their Zeeman-split spin sublevels and their polarized emission. $|0\rangle$ is the crystal ground state.}
\end{center}
\end{figure*}

The dynamics of the $\sigma^+$ and $\sigma^-$ circularly polarized PL after excitation with $\sigma^+$ polarized pulses are shown in Fig.~\ref{fig3}a. The strong difference in their amplitudes in favor of the $\sigma^+$ polarized signal demonstrates a large degree of the optical orientation defined as: 
\begin{equation}
\label{eq3}
P_{\rm{oo}} = \frac{I^{++} - I^{+-}}{I^{++} + I^{+-}}\,,
\end{equation}
and is plotted as a function of time in Fig.~\ref{fig3}b. Here $I^{++}$ and $I^{+-}$ are the intensities of the $\sigma^+$ and $\sigma^-$ polarized emission for $\sigma^+$ polarized excitation. Strikingly, the initial value of $P_{\rm{oo}}$ reaches 0.85 (85\%), dropping during the first 100~ps to a saturation level of 0.60 (60\%) without showing almost no further decay. Commonly, the decay of $P_{\rm{oo}}(t)$ is attributed to spin relaxation. In our case situation is different as both excitons and electron-hole recombination are contributing to $P_{\rm{oo}}$ and exciton spin relaxation time exceeds its lifetime $\tau_\text{X}$ (see below and in SI, S3). At the initial stage excitons mainly contribute to $P_{\rm{oo}}(t)$. After their recombination the PL signal is dominated by long lived carriers with $P_{\rm{oo}} = 0.60$. Therefore, the $P_\text{oo}$ decay from 0.85 to 0.60 is determined by the exciton lifetime. 

 Assuming, in agreement with experiment, that the lifetime and spin relaxation time of electrons and holes exceeds by far the exciton lifetime, we have for the limiting value of the circular polarization degree~\cite{OO_book}
\begin{equation}
P_\text{oo}(t\gg\tau_{\rm X},\tau_{\rm s}) = \frac{\tau_\text{s}}{\tau_{\text{X}} + \tau_\text{s}}P_{\rm oo}(0) \,. 
\label{eq:OO}
\end{equation}
Here $\tau_{\rm s}$ is the exciton spin relaxation time, and $\tau_{\rm X}$ its lifetime. The initial polarization value, $P_{\rm oo}(0)$, is limited by the maximum $P^{\rm max}_\text{oo}=1$ (100\%), dictated by the band structure for transitions at the absorption edge. Taking $\tau_\text{X}=\tau_\text{R1}=55$~ps and $P_\text{oo}=0.85$ from experiment, we obtain $\tau_\text{s}=220$~ps. 

Surprisingly, such a high degree of initial optical orientation is measured for a considerable detuning of the laser energy from the exciton resonance by $E_\text{exc} - E_\text{X}= 0.163$~eV. By contrast, in conventional semiconductors due to: (i) the modification of the selection rules caused by band mixing and (ii) the efficient Dyakonov-Perel mechanism resulting in accelerated spin relaxation of charge carries with large energies, a high $P_\text{oo}$ can be detected only under resonant or close-to-resonant excitation.  

In Fig.~\ref{fig3}c $P_{\rm oo}(0)$ is shown for a large range of excitation energy detunings from 0.08 up to 1.2~eV. The value of 0.85 is preserved up to the detuning of 0.3~eV and then it smoothly decreases, approaching zero at 1.05~eV. All these detunings exceed by far the exciton binding energy so that the photogenerated electron and hole quickly separate in space because of their opposite momenta. Thus, the direct formation of excitons from photogenerated electron-hole pairs is unlikely. Hence, the carriers relax within a few picoseconds to the bottom of the conduction and top of the valence bands and, at cryogenic temperatures, can either form excitons or become localized, where they can recombine as spatially separated carriers. The large value of $P_\text{oo}$ for the excitons indicates  that: (i) the chiral selection rules are fulfilled not only at the absorption edge, but even for large detunings, and (ii) the carriers do not loose their spin polarization during energy relaxation, but show almost negligible spin relaxation within the exciton lifetime at $T=1.6$~K.  

We have checked that the depolarization in a transverse magnetic field via the Hanle effect, Fig.~\ref{fig3}f, gives a similar exciton spin relaxation time of 300~ps (SI, S5C). The huge value of $P_{\rm oo}=0.6$ after exciton recombination, Fig.~\ref{fig3}b, demonstrates a remarkably long free carrier spin relaxation. The estimates based on the Hanle effect model give $\tau_{\rm s, e/h} \approx 1200$~ps (SI, S5C).

We find a strong temperature dependence of $P_\text{oo}$ (Fig.~\ref{fig3}e) with a vanishing polarization for temperatures exceeding 30~K. We attribute that to thermally activated spin relaxation for excitons and free carriers~\cite{kirstein2021}.

\begin{figure*}[hbt]
\begin{center}
\includegraphics[width = 14cm]{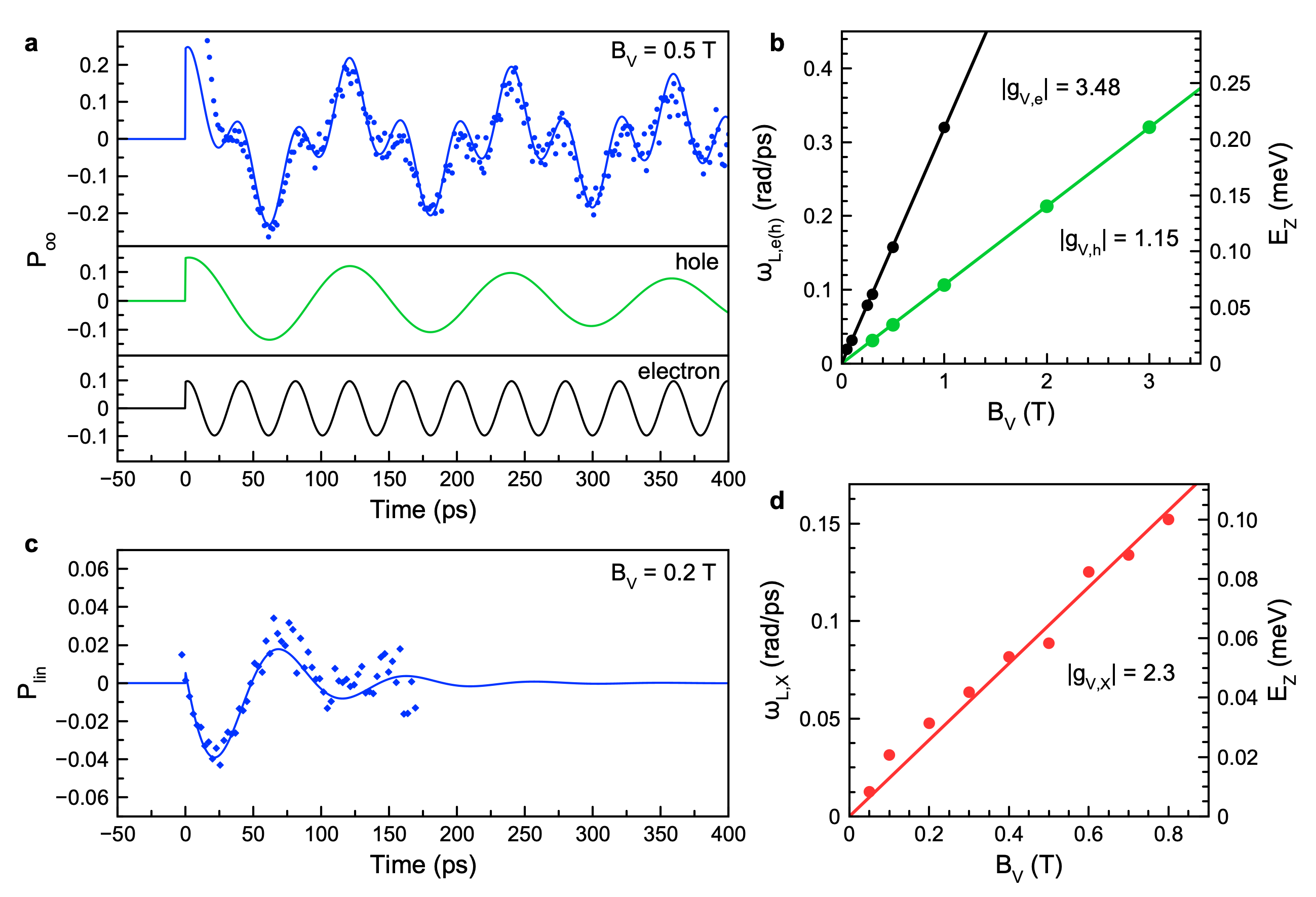}
\caption{\label{fig4} Spin precession of excitons and resident carriers in a Voigt magnetic field measured by TRPL at $T = 1.6$~K.
(a) Dynamics of the optical orientation degree $P_{\rm oo}(t)$ measured for $B_\text{V} = 0.5$\,T using $\sigma^+$ excitation at $E_\text{exc} = 1.675$\,eV with $P = 30$\,mW/cm$^2$ (symbols). $E_\text{det} = 1.506$\,eV. Blue line is a fit with Eq.~\eqref{Poo_beats} including electron and hole contributions, which are shown by the black and green lines, respectively.
(b) Dependences of the Larmor precession frequencies of electron (black circles) and hole (green circles) on $B_\text{V}$. Linear fits provide $|g_\text{V,e}| = 3.48$ and $|g_\text{V,h}| = 1.15$.
(c) Dynamics of the linear polarization degree $P_\text{lin}(t)$ measured at $B_\text{V} = 0.2$\,T using $\sigma^+$ polarized excitation as in panel (a) (the symbols). Line is a fit with Eq.~\eqref{Plin_beats} using $\omega_{\rm{L,X}}=0.048$~rad/ps and $\tau_{\rm{X}}=55$~ps. (d) Magnetic field dependence of the Larmor precession frequency from $P_\text{lin}(t)$ (the symbols). Line is a linear fit with $|g_\text{V,X}| = 2.3$. Right scale gives the corresponding Zeeman splittings. 
}
\end{center}
\end{figure*}

\subsection*{Polarization of bright excitons in longitudinal magnetic field}


In order to address the spin dynamics of the excitons in their ground state and separate it from the spin dynamics of carriers at larger energies, we measure and analyze the degree of circular polarization (DCP) induced by a magnetic field applied in the Faraday geometry, $P_\text{c}(B_\text{F})$, using linearly polarized excitation. Polarized PL spectra integrated over exciton lifetime are shown in Fig.~\ref{fig2}a. The exciton line at 1.506~eV demonstrates the Zeeman splitting of the bright (optically-active) exciton states with angular momentum $z$-components $J_{z} =\pm 1$: $E_\text{Z} = \mu_\text{B} g_\text{X} B$, where $g_\text{X}$ is the exciton $g$-factor and $\mu_\text{B}$ is the Bohr magneton.  The magnetic field dependence of the Zeeman splitting (Fig.~\ref{fig2}b) gives us $g_\text{F,X} = +2.4$, which coincides well with the known values for FA$_{0.9}$Cs$_{0.1}$PbI$_{2.8}$Br$_{0.2}$~\cite{Kopteva_gX_2023}. Positive sign of $g_{\rm X}$ corresponds to the lower in energy $\sigma^-$ polarized sublevel, see inset in Fig.~\ref{fig2}c.

Figure~\ref{fig2}a shows that the PL emission becomes circularly polarized in magnetic field with stronger emission in $\sigma^-$ polarization for $B_\text{F}>0$, evidencing a stronger population of the lower energy Zeeman sublevel, see inset of Fig.~\ref{fig2}c. The DCP is evaluated as~\cite{OO_book}
\begin{equation}
\label{eq1}
P_\text{c}(B_\text{F}) = \frac{I^+ - I^-}{I^+ + I^-},
\end{equation}
where $I^+$ and $I^-$ are the intensities of the $\sigma^+$ and $\sigma^-$ polarized emission components. The $P_\text{c}$ dependence on $B_\text{F}$ is shown in Fig.~\ref{fig2}c. The DCP magnitude increases linearly in small magnetic fields and approaches saturation at $P_\text{c} = -0.20$ at $B_\text{F} = 5$~T.  Such behavior is typical for excitons thermalized on the Zeeman levels and can be described by:
\begin{equation}
\label{eq2}
P_\text{c}(B_\text{F}) = \frac{\tau_{\text{X}}}{\tau_{\text{X}} + \tau_{\text{s}}}\tanh \left(-\frac{g_\text{F,X}\mu_\text{B}B_\text{F}}{2k_\text{B}T} \right).
\end{equation}
We fit the experimental data with this equation using $T = 1.6$~K  and $\tau_{\text{X}} = 55$~ps, see the red line in Fig.~\ref{fig2}c. It gives $\tau_{\text{s}} = 4.6\tau_{\text{X}}=250$~ps as the only fit parameter, which is close to the $\tau_{\text{s}}=220$~ps extracted from optical orientation. Interestingly, the long spin relaxation time compared to the lifetime results in opposite trends of DCP and optical orientation: (i) it significantly reduces the DCP, as compared to the fully thermalized case ($\tau_{\rm s} \ll \tau_{\rm X}$) shown by the blue line in Fig.~3c, and (ii) it results in high values of the optical orientation, see Eq.~\eqref{eq:OO}.

\begin{figure*}[hbt]
\begin{center}
\includegraphics[width = 16cm]{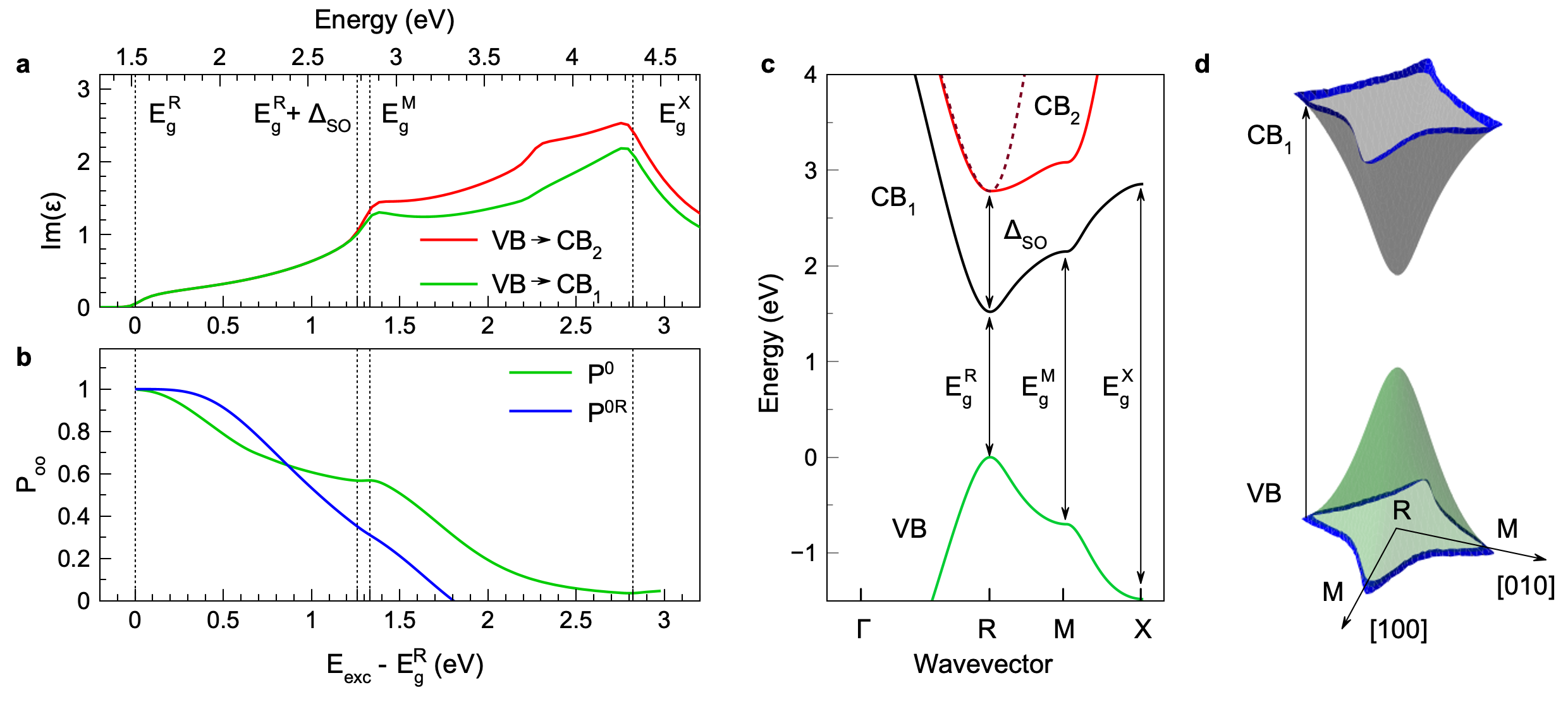}
\caption{\label{fig5} Calculation of optical orientation in lead halide perovskites. (a) Imaginary part of the dielectric function calculated in the empirical tight-binding method for the VB$\rightarrow$CB$_1$ (green) and VB$\rightarrow$CB$_2$ (red) transitions, using the parameters from Ref.~\citenum{Nestoklon2021} corrected for experimental data, see SI, S6. Vertical lines show the energies of the transitions at the $R$, $M$, and $X$ points. The upper axis is the energy detuning from the band gap energy at the $R$ point ($E_\text{g}^\text{R}$). 
(b) Dependence of the optical orientation degree on the excitation energy detuning from $E_\text{g}^\text{R}$, $E_\text{exc}-E_\text{g}^\text{R}$, calculated within the two models explained in the text. Vertical lines show the energies of the $R$, $M$, and $X$  points.
(c) Band diagram of the bulk {FA}$_{0.9}$Cs$_{0.1}$PbI$_{2.8}$Br$_{0.2}$ crystal along the $\Gamma \to  R \to M \to  X$ path.  
(d) Illustration of the wave vectors ${\bf k}_0$ in the (001) plane contributing to the transitions with energy close to $E_{\rm g}^{\rm M}$. One can see the large warping of the band structure leading to electron/hole distributions highly anisotropic in the $k$-space.}
\end{center}
\end{figure*}

\subsection*{Spin precession in transverse magnetic field}

The information of investigating the spin dynamics can be enriched by application of a magnetic field in the Voigt geometry, perpendicular to the light $k$-vector ($\textbf{B}_{\rm V} \perp \textbf{k}$). In this case, the exciton and/or carrier spins, which have been optically oriented along the $k$-vector, undergo Larmor precession about the field direction with the frequency $\omega_{\rm L}=g\mu_{\rm B}B_{\rm V}$. The spin dynamics measured thereby provide access to the $g$-factor value and to spin dephasing time $T^*_2$, the latter from the signal decay. TRPL\cite{Heberle1994,Dyakonov1997,Marie2000} can be used to directly monitor the spin precession quantum beats of excitons and charge carriers via the circular and linear polarization degree of emission.

Figure~\ref{fig4}a shows the dynamics of $P_{\rm oo}(t)$ measured for $B_\text{V} = 0.5$~T at the exciton energy. A complex pattern of spin beats is observed with a very weak decay within the temporal range of 400~ps. The corresponding decay time greatly exceeds the exciton lifetime, so that we assign the signal to the coherent spin precession of resident carriers. The signal contains two oscillating components with Larmor frequencies corresponding to the $g$-factors of the electron ($|g_\text{V,e}| = 3.48$) and the hole ($|g_\text{V,h}| = 1.15$), see the fits in Fig.~\ref{fig4}a and the magnetic field dependencies of the Larmor frequencies in Fig.~\ref{fig4}b, in agreement with the time-resolved Faraday/Kerr rotation measurements on the same perovskite crystal~\cite{kirstein2021}. The absence of any offset in the Zeeman splittings for $B_{\rm V}\to 0$ confirms that the signal arises from pairs of spatially-separated electrons and holes whose exchange interaction is negligible. The dependence $P_{\rm oo}(t)$ is accurately described by the model approach developed in SI, S5 for the case with $\Delta_{\rm X}=0$, Fig.~S4b.

Strikingly, in the Voigt geometry the exciton PL for circularly polarized excitation becomes linearly polarized with a degree defined as $P_\text{lin} = (I^{+\perp} - I^{+\parallel})/(I^{+\perp} +I^{+\parallel})$. Here, $I^{+\perp}$ and $I^{+\parallel}$ are the PL intensities  in linear polarizations perpendicular and paraller to the  magnetic field direction. The dynamics of $P_\text{lin}(t)$ measured at $B_\text{V} = 0.2$~T are shown in Fig.~\ref{fig4}c. The polarization degree decays with the time $\tau_\text{X} = 55$~ps, during which it precesses with the Larmor frequency, corresponding to $|g_\text{V,X}| = 2.3$ (Fig.~\ref{fig4}d), close to the exciton $g$-factor measured from PL $g_\text{F,X} = +2.4$ (Fig.~\ref{fig2}b) and to the sum of the carrier $g$-factors $g_\text{V,e} + g_\text{V,h} = +2.33$. These facts allow us to reliably assign the spin beats detected in the linear polarization degree to the dynamics of the bright exciton states with $J_{z} = \pm 1$ with a finite exchange interaction ($\Delta_\text{X} >0$), see SI, S5A. These linear polarization beats are a clear indication of electron and hole spin correlations. The analysis in SI, S5B shows that the $P_{\rm lin}$ is governed by the quantum mechanical average $\langle \hat{s}_{x}^{\rm e} \hat{s}_{x}^{\rm h} - \hat{s}_{y}^{\rm e} \hat{s}_{y}^{\rm h}\rangle$ with $\bm \hat{s}^{\rm e/h}$ being the electron and hole spin operators and $x,y$ labeling their in-plane components. Hence, the polarization of both electron and hole spins is needed to obtain linear polarization, in stark contrast to the case of circular polarization which can appear as result of recombination of a polarized carrier with an unpolarized one. 

We stress that the beats both in circular and linear polarization can be excited highly-nonresonantly, for example, with the detuning $E_\text{exc} - E_\text{X} = 0.17$\,eV.

\subsection*{Theoretical analysis}
\label{sec:theory}

Bulk perovskites are strongly different from conventional III-V and II-VI semiconductors~\cite{OO_book,Spin_book_2017}, owing to the presence of an inversion center in the point symmetry group and an ``inverted'' band structure with the simple, two-fold degenerate conduction and valence bands at the $R$ point of the Brillouin zone, Fig.~\ref{fig3}d. The former nullifies the spin-orbit splitting of the bands and suppresses the spin relaxation of charge carriers due to  absence of the Dyakonov-Perel' spin relaxation mechanism. The latter allows for 100\% optical orientation of charge carriers and excitons at the interband transitions. The measured high degree of optical orientation and its robustness against excitation detuning are in agreement with these qualitative considerations. However, the loss of optical orientation for excitation with a detuning $E_\text{exc} - E_\text{X}  > 0.6$~eV (Fig.~\ref{fig3}c) calls for a detailed theoretical analysis.

The general scheme of optical orientation calculations includes three basic steps: (1) A circularly polarized optical pulse generates electrons and holes with the wave vector ${\bf k}_0$ satisfying the momentum and energy conservation laws: the momentum is equal for both carriers and their energy is equal to the incident photon energy (here we neglect the momentum of incident photon). For a fixed photon energy, there is a set of possible ${\bf k}_0$. The spin states of the photoexcited carriers are defined by the polarization of the incident light and can be described by the density matrix of the carriers $\hat{\rho}^{c,v}({\bf k}_0)$. (2) The electrons and holes loose their kinetic energy, mostly due to interaction with optical phonons. Assuming a definite mechanism of energy relaxation, one may compute the density matrix of the carriers at the $R$ point $\hat{\rho}^{c,v}({\bf k}_R)$. (3) The electrons and holes recombine at the $R$ point.

We start theoretical analysis with calculating the absorption. We use the tight-binding approach with the parameters of CsPbI$_3$ from Ref.~\citenum{Nestoklon2021} corrected for the measured band gap, see SI, S6. The imaginary part of the dielectric function without excitonic effects is presented in Fig.~\ref{fig5}a.  In agreement with previous calculations\cite{BoyerRichard2016,Song2020}, the absorption increases from the band gap at the $R$ point up to the $M$ point, see Fig.~\ref{fig5}c for the band dispersion.

Figure~\ref{fig5}b shows the optical orientation degree calculated in the two models described in SI, S6: $P^{\rm 0R}$ corresponds to the ``effective phonon'' model, where the direct transition from $\mathbf k_0$ to the $R$ point is assumed to occur via the spin-independent electron/hole-phonon interaction, and $P^{0}$ corresponds to the ``effective emission'' model where we calculate the polarization of the emission assuming recombination in the excited states. The band energies, matrix elements of velocity, and amplitudes of the transitions between ${\bf k}_0$ and ${\bf k}_R$ are calculated in the empirical tight-binding approach (SI, S6). As can be seen in Figs.~\ref{fig5}d  and \ref{fig3}c, the calculated $P_{\rm oo}(E_{\rm exc})$ dependence qualitatively follows the experimental data: it decreases starting from the detuning of about 0.3~eV, but remains large $>0.4$ up to detunings of $\sim 1.0$~eV. From the behavior of $P^0(E_{\rm exc})$ we conclude that the depolarization to a large extent originates from deviations of the selection rules from the strict ones away from the $R$ point. For even higher detunings $\gtrsim 2.8$~eV, the transitions to the spin-split-off conduction band, CB$_2$, set in which further reduce the polarization (not included in calculations). Additional depolarization results from spin-flip scattering of charge carriers during their energy relaxation (SI, S6).
  
\section*{conclusions}

In conclusion, we use time- and polarization-resolved photoluminescence to demonstate a giant, unprecedentedly high optical orientation degree up to $85\%$ in bulk perovskite crystals, which is amazingly robust with respect to a significant, up to $\sim 0.3$~eV, detuning from the fundamental absorption edge. The combination of symmetry analysis and atomistic calculations shows that the remarkable optical orientation is a consequence of the unique properties of the lead halide perovskites: ``clean'' chiral selection rules for the optical transitions between the two-fold degenerate valence and conduction bands and the suppressed spin relaxation owing to the absence of a bands' spin splitting, resulting from the presence of the inversion center. In time-resolved circularly polarized luminescence for non-resonant excitation in a transverse magnetic field, we observe coherent spin precession of electrons and holes, providing direct access to their Land\'e factors corroborating that the spin orientation of the charge carriers is preserved during their course of energy relaxation. Importantly, we observe for the same conditions linear polarization of the emission which serves as unequivocal indication of electron and hole spin correlations in the perovskites. Combined with the simple fabrication and the bright optical properties, these features make lead halide perovskites a prime material system for spintronic technologies.  

\section*{Methods}

\textbf{Samples.}  
The studied {FA}$_{0.9}$Cs$_{0.1}$PbI$_{2.8}$Br$_{0.2}$ bulk single crystal was grown out of solution with the inverse temperature crystallization technique.\cite{nazarenko2017} A solution of CsI, FAI, PbI$_{2}$ and PbBr$_{2}$ is mixed with GBL $\gamma$-butyrolactone solvent. The solution is filtered and heated to $130^\circ$C, so the crystals are formed in the $\alpha$-phase.\cite{weller2015} Single crystals are separated by filtration and drying. The $\alpha$-phase or black phase of {FA}$_{0.9}$Cs$_{0.1}$PbI$_{2.8}$Br$_{0.2}$ has a cubic crystal structure at room temperature. Further growth details are given in the Supplementary Note~1. Since the crystal shows a $g$-factor isotropy at low temperatures, we consider its structure also then as cubic.\cite{kirstein2021} The geometry with light wave vector $\textbf{k}\parallel [001]$ was used in all optical experiments.

\textbf{Magneto-optical measurements.} 
For low-temperature optical measurements we use a liquid helium cryostat with the temperature variable from 1.6\,K up to 300\,K. At $T=1.6$\,K the sample is placed in superfluid helium, while at $4.2-30$\,K it is held in helium vapor. A superconducting magnet equipped with a pair of split coils can generate a magnetic field up to 5\,T. The cryostat is rotated by $90^\circ$ to change the experimental geometry: The magnetic field parallel to $\textbf{k}$ is denoted as $B_{\rm F}$ (Faraday geometry) and perpendicular to $\textbf{k}$ as $B_{\rm V}$ (Voigt geometry).

\textbf{Photoluminescence and photoluminescence excitation measurements.} 
The time-integrated photoluminescence spectrum (PL) was measured with a 0.5\,m spectrometer equipped with a charge-coupled-devices (CCD) camera. For PLE, the PL intensity at the energy $E_{\rm{det}} = 1.496$\,eV was detected as function of the excitation energy of a tunable titanium-sapphire continuous wave laser. 

\textbf{Time-resolved photoluminescence.}
The spectrally resolved PL dynamics were measured using a spectrometer with a 300 grooves/mm diffraction grating and a streak camera with 10~ps time resolution. Pulses with 200\,fs duration and photon energies from 1.59\,eV (780\,nm) to 2.67\,eV (465\,nm) from a tunable Chameleon Discovery laser with the repetition rate of 80~MHz were used for PL excitation. The time-integrated PL spectrum is obtained from time integration of the PL dynamics. To study the effect of optical orientation, we used circularly ($\sigma^+/\sigma^-$) polarized excitation light and analyzed the circularly or linearly polarized emission.

The dynamics of the full intensity (proportional to the population) show a decay after pulse action with multiple recombination times ($\tau_{\rm{Ri}}$):
\begin{equation}
I(t) = \sum_{j = 1,2}I_i(0)\exp(-t/\tau_{\rm{Rj}}),
\end{equation}
where $I_i(0)$ is the initial population of each component.

The dynamics of the optical orientation degree can be described by a decaying oscillatory function:
\begin{equation}
\label{Poo_beats}
P_{\rm{oo}}(t) = \sum_{i}P_{\rm{oo}}(0) \cos(\omega_{\rm{L,i}}t)\exp(-t/\tau_{\rm{s,i}}). 
\end{equation}
Here $P_{\rm{oo}}(0)$ is the spin polarization degree at zero time delay, the index $i = $e,h denotes the electron or hole component to the Larmor precession frequency $\omega_{\rm{L},i}$ and in the spin relaxation time $\tau_{\rm{s},i}$. The exciton Larmor precession in the degree of linear polarization is described by:
\begin{equation}
\label{Plin_beats}
P_{\rm{lin}}(t) = \sum_{i}P_{\rm{lin}}(t=0) \cos(\omega_{\rm{L,X}}t)\exp(-t/\tau_{\rm{X}}).
\end{equation}

\textbf{Theoretical analysis. }
The band energies, matrix elements of velocity, and amplitudes of the optical transitions between ${\bf k}_0$ and ${\bf k}_R$ are calculated in the empirical tight-binding approach. We use the tight-binding approach with the parameters of CsPbI$_3$ from Ref.~\citenum{Nestoklon2021}. The matrix elements of velocity are calculated following Ref.~\citenum{Graf1995}. In calculations we take a $50\times50\times50$ $k$-mesh in $1/8$ of the (cubic) Brillouin Zone. For more details see SI, S6.

\textbf{Data availability.} 
The data on which the plots in this paper are based and other findings of our study are available from the corresponding authors upon justified request. \\

\subsection*{Acknowledgements}
The authors are thankful to  Al. L. Efros, I. A. Yugova and V. L. Korenev for fruitful discussions. We acknowledge the financial support by the Deutsche Forschungsgemeinschaft via the SPP2196 Priority Program (Projects YA 65/26-1 and AK 40/13-1). The work at ETH Z\"urich (O.H., D.N.D. and M.V.K.) was financially supported by the Swiss National Science Foundation (grant agreement 186406, funded in conjunction with SPP219 through the DFG-SNSF bilateral program) and by ETH Z\"urich through ETH+ Project SynMatLab.

\subsection*{Author contributions}
D.R.Y., I.A.A., and N.E.K. conceived the experiment. E.Y., N.E.K., M.K., D.K., E.K., and E.A.Z. built the experimental apparatus and performed the measurements. E.Y., N.E.K., D.R.Y., M.K., E.K., E.A.Z., and I.A.A.  analyzed the data. 
M.O.N., M.M.G., and N.E.K. provided the theoretical description.  
O.H., D.N.D., and  M.V.K. fabricated and characterized the samples. All authors contributed to interpretation of the data. N.E.K., D.R.Y., M.O.N., M.M.G., and I.A.A. wrote the manuscript in close consultation with M.B.

\subsection*{Additional information}
Correspondence should be addressed to N.E.K. (email: natalia.kopteva@tu-dortmund.de) and D.R.Y. (email: dmitri.yakovlev@tu-dortmund.de)

\subsection*{Competing financial interests}
The authors declare no competing financial interests.

\subsection*{ORCID} 
Natalia E. Kopteva:   0000-0003-0865-0393 \\
Dmitri R. Yakovlev:   0000-0001-7349-2745 \\
Ey\"up Yalcin:    0000-0003-2891-4173  \\
Ilya A. Akimov:   0000-0002-2035-2324    \\
Evgeny~A.~Zhukov: 0000-0003-0695-0093   \\ 
Erik~Kirstein:    0000-0002-2549-2115 \\
Mikhail O. Nestoklon: 0000-0002-0454-342X\\
Mikhail M. Glazov:    0000-0003-4462-0749 \\ 
Mladen Kotur: 0000-0002-2569-5051\\
Dennis Kudlacik: 0000-0001-5473-8383\\
Oleh Hordiichuk: 0000-0001-7679-4423\\
Dmitry N. Dirin: 	0000-0002-5187-4555 \\
Maksym~V.~Kovalenko:  0000-0002-6396-8938 \\
Manfred Bayer:        0000-0002-0893-5949\\

\end{bibunit}
\clearpage
\newpage
\begin{mywidetext}
\begin{center}
  \textbf{{\Large Supplementary Information:}}\\
  \textbf{{\Large Giant optical orientation of exciton spins in lead halide perovskite crystals}}
\end{center}
\end{mywidetext}
\setcounter{equation}{0}
\setcounter{figure}{0}
\setcounter{table}{0}
\setcounter{page}{1}
\setcounter{section}{0}
\makeatletter
\renewcommand{\thepage}{S\arabic{page}}
\renewcommand{\theequation}{S\arabic{equation}}
\renewcommand{\thefigure}{S\arabic{figure}}
\renewcommand{\thetable}{S\arabic{table}}
\renewcommand{\thesection}{S\arabic{section}}
\renewcommand{\bibnumfmt}[1]{[S#1]}
\renewcommand{\citenumfont}[1]{S#1}


\renewcommand{\S}{\mathop{\mathcal S}}
\newcommand{\X}{\mathop{\mathcal X}}
\newcommand{\Y}{\mathop{\mathcal Y}}
\newcommand{\Z}{\mathop{\mathcal Z}}
\begin{bibunit}

\onecolumngrid

\section{Material information}

The studied lead halide perovskite crystals belong to the FAPbI$_{3}$ material class. FA-based perovskites exhibit a low trap density ($1.13 \times 10^{10}$~cm$^{-3}$) and a low dark carrier density ($3.9 \times 10^{9}$~cm$^{-3}$)~\cite{nazarenko2017,zhumekenov2016}. FAPbI$_3$ is chemically and thermally more stable compared to MAPbI$_3$ due to the possible decomposition of the latter to gaseous hydrogen iodide and methylammonium~\cite{nazarenko2017}. However, pure FAPbI$_3$ suffers from structural instability originating from the large size of FA the cation which cannot be accommodated by the inorganic perovskite framework. This instability has been successfully resolved via partial, up to 15\%, replacement of the large FA cation with the smaller caesium (Cs) together with the iodine (I) partial substitution by bromide (Br) \cite{mcmeekin2016_SI,jeon2015_SI}. As a result, the Goldschmidt tolerance factor~\cite{goldschmidt_gesetze_1926} $t$ is tuned from 1.07 in FAPbI$_3$ closer to 1 in FA$_{0.9}$Cs$_{0.1}$PbI$_{2.8}$Br$_{0.2}$, where it is 0.98. The band gap of FA$_{0.9}$Cs$_{0.1}$PbI$_{2.8}$Br$_{0.2}$ at room temperature is 1.52~eV, slightly larger than the band gap of FAPbI$_{3}$ of 1.42~eV~\cite{li2016,nazarenko2017}. 

For crystal synthesis the inverse temperature crystallization technique is used~\cite{nazarenko2017,zhumekenov2016}. For the growth, a solution of CsI, FAI (FA being formamidinium), PbI$_2$, and PbBr$_2$, with GBL $\gamma$-butyrolactone as solvent is mixed. This solution is then filtered and slowly heated to 130$^\circ$C temperature, whereby single crystals are formed in the black phase of FA$_{0.9}$Cs$_{0.1}$PbI$_{2.8}$Br$_{0.2}$, following the reaction
\begin{equation*}
\mathrm{PbI}_2 + \mathrm{PbBr}_2 + \mathrm{FAI} + \mathrm{CsI} \underset{}{\stackrel{[GBL]}{\rightarrow}} \mathrm{FA}_{0.9}\mathrm{Cs}_{0.1}\mathrm{PbI}_{2.8}\mathrm{Br}_{0.2} + \mathrm{R}.
\end{equation*}
Afterwards, the crystals are separated by filtering and drying. A typical crystal as used for this study has a size of about 2~mm. The crystallographic analysis suggests that one of the principal axes $a$, $b$, $c$ is normal to the front facet, thus pointing along the optical axis. In cubic approximation $a=b=c$. The pseudo-cubic lattice constant for hybrid organic perovskites (HOP) is around 0.63~nm~\cite{whitfield2016}, but was not determined for this specific sample.

\section{Time-resolved photoluminescence in the microsecond range}

The time-resolved photoluminescence (TRPL) detection technique with high temporal resolution of 10~ps by means of a streak-camera covers a time range limited to 2~ns. In order to measure longer times, we use pump-probe differential reflectivity or TRPL using a time-of-flight computer board. In FA$_{0.9}$Cs$_{0.1}$PbI$_{2.8}$Br$_{0.2}$ at $T=6$~K the decay of the exciton population dynamics measured by pump-probe differential reflectivity occurs with a time constant of 450~ps~\cite{kirstein2021}. This is in good agreement with the $\tau_\text{R2} = 380 - 840$~ps obtained from streak camera measurements, see Figure~1c. However, the PL dynamics contain also long decay times, which are not related to exciton recombination.

The long-lived recombination dynamics up to 100~$\mu$s are measured by the TRPL technique. Here, the PL is excited by a pulsed laser with parameters: photon energy of 2.33~eV (wavelength of 532~nm), pulse duration of 5~ns, repetition rate of 10~kHz, and average excitation power of 8~$\mu$W. The detection energy was selected by a Jobin-Yvon U1000 double monochromator (1 meter focal length). The signal was detected using an avalanche photodiode connected to time-of-flight card with time resolution of 30~ns. 

\begin{figure*}[hbt]
\begin{center}
\includegraphics[width = 18cm]{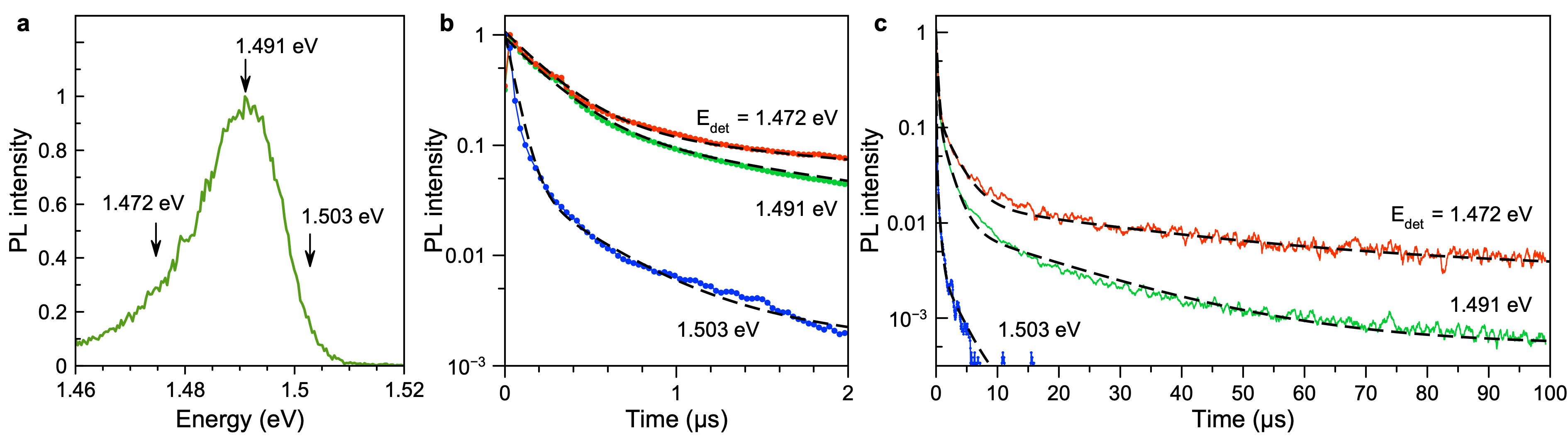}
\caption{\label{figS1} (a) Time-integrated PL measured for pulsed excitation with $E_\text{exc} = 2.33$\,eV and repetition rate of  10~kHz. $T = 1.6$~K. Time-resolved PL measured at various energies in 2~$\mu$s range (b) and in 100~$\mu$s range (c). PL dynamics fitted with a multi-exponential function with decay times given in Table~\ref{tab:St1}. The fits are shown by black dashed lines. 
}
\end{center}
\end{figure*}

\begin{table*}[hbt]
\caption{Recombination times for FA$_{0.9}$Cs$_{0.1}$PbI$_{2.8}$Br$_{0.2}$, measured at different spectral positions for $T = 1.6$\,K.}
\label{tab:St1}
\begin{center}
\begin{tabular*}
{0.55\textwidth}{@{\extracolsep{\fill}} |>{\centering\arraybackslash} m{0.1\textwidth}|>{\centering\arraybackslash} m{0.1\textwidth}|>{\centering\arraybackslash} m{0.1\textwidth}|>{\centering\arraybackslash} m{0.1\textwidth}|}
\hline
$E_\text{det}$\,(eV) & $\tau_1$\,(ns)& $\tau_2$\,($\mu$s)& $\tau_3$\,($\mu$s)\\
\hline
1.503 & 370 & 3.0 & -\\
\hline
1.491 & 250 & 1.6 & 20 \\
\hline
1.472 & 200 & 3.3 & 44\\
\hline
\end{tabular*}
\end{center}
\end{table*}

Figure~\ref{figS1} shows the recombination dynamics in a FA$_{0.9}$Cs$_{0.1}$PbI$_{2.8}$Br$_{0.2}$ crystal, detected at various spectral energies. The  time-integrated PL spectrum measured for pulsed laser excitation is shown in Figure~\ref{figS1}a. The arrows indicate the detection energy of the photoluminescence dynamics. The experiments were performed at the temperature of $T = 1.6$\,K. The recombination dynamics cover a broad temporal range up to 100~$\mu$s. They cannot be fitted by a monoexponential decay evidencing that several recombination processes are involved. The PL dynamics comprise three exponential decays in addition to the short exciton recombination decay, which is too fast to be resolved in this experiment, see Figures~\ref{figS1}b,c. The long recombination dynamics can be associated with the following processes: (i) Recombination of electrons and holes localized at different crystal sites with a significant dispersion in their separation lengths~\cite{belykh2019,kirstein2022mapi,kirstein2021,kirstein2021nc}, (ii) Carrier trapping and detrapping processes~\cite{Chirvony2018}, (iii) Polaron formation~\cite{deQuilettes2019}, (iv) Slow carrier diffusion from the sample surface to the crystal depth~\cite{Bercegol2018}. The clarification of the role of the specific mechanisms goes beyond the scope of the present study.

\section{Optical orientation of excitons measured for resonant excitation}

In order to study the direct optical orientation of excitons, we use resonant excitation in pump-probe differential reflectivity. We use a laser system generating pulses  with 1.5~ps duration and repetition rate of 75~MHz. The pump and probe photon energies were the same and were set to the exciton resonance at 1.503~eV. The pump was $\sigma^+$ circularly polarized, its intensity was modulated by an electro-optical modulator at a frequency of 100~kHz. The probe beam was either $\sigma^+$ or $\sigma^-$ circularly polarized. It was reflected from the sample and sent on a photodiode, which output was analyzed by a lock-in amplifier. The temporal resolution was about 2~ps. Note that this technique has a higher time resolution than TRPL. Also both the population dynamics and spin dynamics of excitons and photogenerated carriers can be measured.   
 
\begin{figure*}[hbt]
\begin{center}
\includegraphics[width = 18cm]{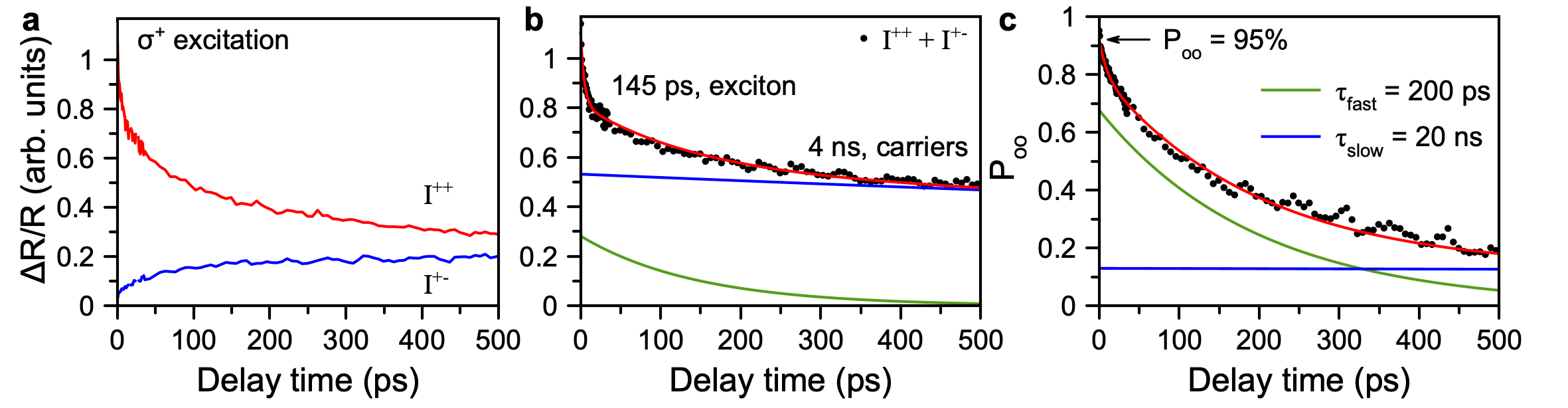}
\caption{\label{figSX} Spin dynamics measured by polarized pump-probe reflectivity at $T = 1.6$~K. (a) $\Delta R/R$ dynamics measured for $\sigma^+$ polarized pump, with the detection probe pulses either $\sigma^+$ (red line) or $\sigma^-$ (blue line) polarized. The laser photon energy $E_\text{exc} = E_\text{det} = 1.503$~eV matches the exciton resonance. The pump power is 15~mW and the probe power is 0.6~mW. (b) The calculated full intensity ($I^{++} + I^{+-}$) shown by the black dots decays with the recombination time of the exciton $\tau_\text{X} = 150$~ps (green) and of spatially separated carriers  $\tau_{\text{R},2} = 4$~ns (blue). The red line is the sum fit including the exciton and resident carriers contributions. (c) Dynamics of the optical orientation $P_\text{oo}(t)$ (dots) calculated from the data from panel (a) with Eq.~(1). the red line is a bi-exponential fit. Green and blue lines show the fast and slow components of the $P_\text{oo}(t)$ dynamics.}
\end{center}
\end{figure*}

Figure~\ref{figSX}a shows the dynamics of the differential reflectivity ($\Delta R/R$) signal detected in $\sigma^+$ polarization (red line, denoted as $I^{++}$) and $\sigma^-$ polarization  (blue line, $I^{+-}$) polarization of the probe. $I^{++}$ decays in time, but $I^{+-}$ rises. From a double-exponential fit of the full intensity ($I^{++} + I^{+-}$) shown in Figure~\ref{figSX}b, we obtain a fast decay time of $150$~ps and a slow decay time of $4$~ns. The fast time can be assigned to the exciton lifetime, and the slow time to the recombination of spatially separated resident electrons and holes. the dynamics of the optical orientation degree, $P_\text{oo}(t)$, are shown in Figure~\ref{figSX}c. Its maximal value at a time delay of a few picoseconds reaches 0.95 (i.e. 95\%), which in fact is close to the maximal possible value of 100\%. Note, that we do not show or use in our analysis the strong signal during the time overlap of the pump and probe pulses. 

\section{Effect of excitation density}


Here we present experimental results on the modification of the recombination and spin dynamics of excitons and resident carriers with increasing excitation density. It allows us to better understand the interplay of the exciton and carrier contributions to the emission at the same spectral energy of 1.506~eV. We show below that in this case the dynamics of the optical orientation degree is determined not by the spin relaxations time, as it is common when only one recombination process is involved, but rather by the fast recombination of excitons. Therefore one has to be careful with interpretations. Note that this situation is rather unusual for the spin physics of semiconductors, and we are not aware of a corresponding discussion in literature.

\begin{figure*}[hbt]
\begin{center}
\includegraphics[width = 12cm]{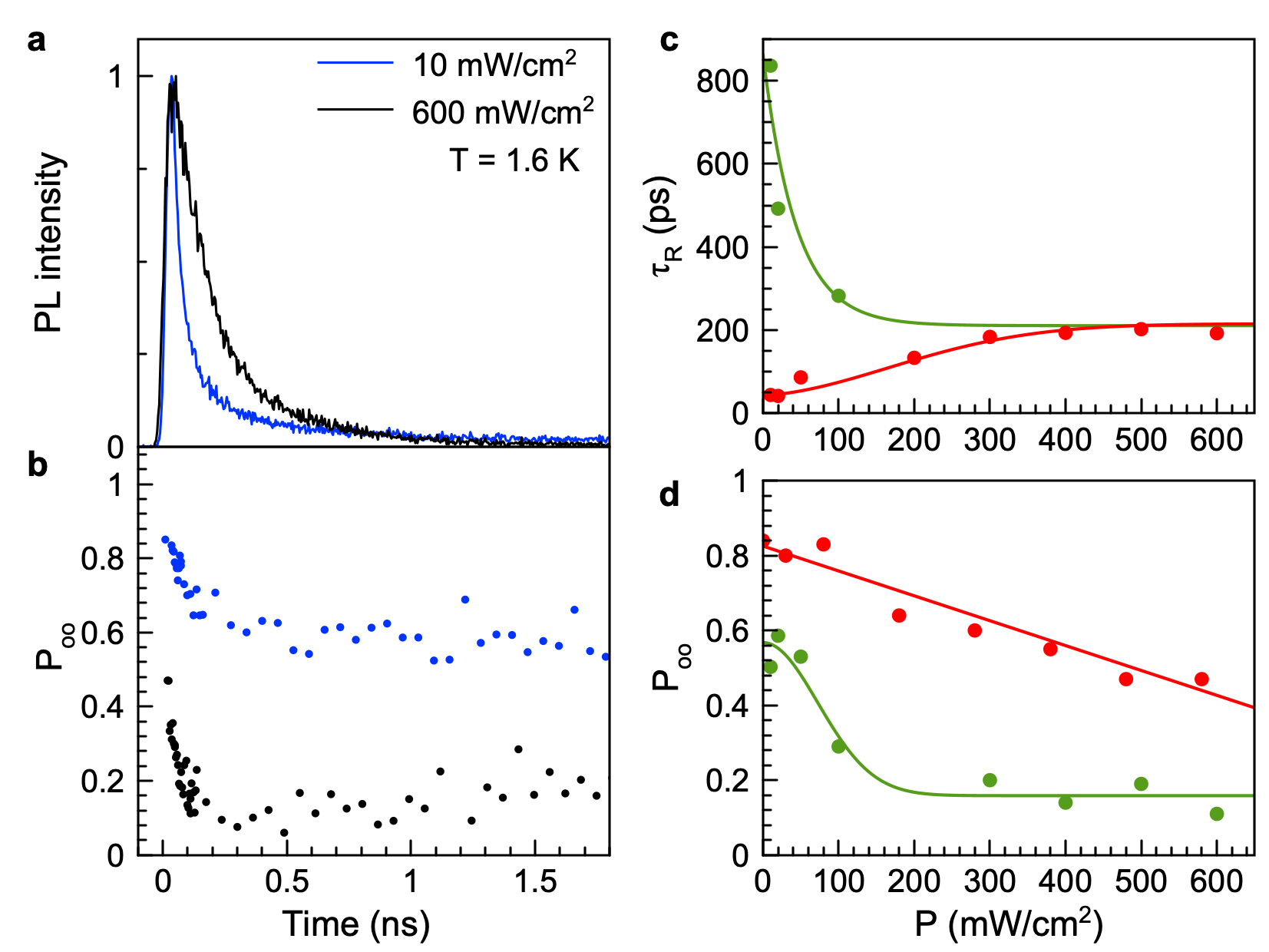}
\caption{\label{figS2} (a) TRPL signals for $P = 10$\,mW/cm$^2$ (blue line) and for $P = 600$\,mW/cm$^2$  (black line). $T = 1.6K$, $E_\text{det} = 1.506$~eV, $E_\text{exc} = 1.675$~eV. (b) Time-resolved $P_\text{oo}$ for $P = 10$\,mW/cm$^2$ (blue dots) and for $P = 600$\,mW/cm$^2$ (black dots). (c) $\tau_{\text{R}1}$ (red dots) and $\tau_{\text{R}2}$ (green dots) dependences on the excitation density. Lines are guides to the eye. (d) Optical orientation of excitons (measured at $t = 0$~ns, red dots) and of localized carriers (measured at $t = 1.5$\,ns, green dots) as function of excitation density. Lines are guides to the eye.} 
\end{center}
\end{figure*}

In the main text, we present experimental data measured at small excitation densities of 10 to 30~mW/cm$^2$. The excitation density of 10~mW/cm$^2$ corresponds to the exciton density of $10^{13}$~cm$^{-3}$. The density of photogenerated electron-hole pairs is low, so that the exciton-exciton interaction can be neglected. Here we show  results for recombination and $P_{\rm oo}(t)$ dynamics, measured at excitation densities varied across a wide range from 10 up to 600~mW/cm$^2$.  Figure~\ref{figS2}a compares the recombination dynamics at 10~mW/cm$^2$ (black line) and 600~mW/cm$^2$ (blue line). One can see that with increasing the excitation density the initial decay becomes slower, the decay time increases from 55~ps up to 200~ps. By contrast, the slow contribution to the dynamics becomes faster, the corresponding decay time shortens from 840~ps down to 200~ps. A redistribution of intensity between the fast and slow contributions also takes place. As a result, the initial decay of PL intensity by about 95\% becomes monoexponential at $P=600$~mW/cm$^2$ with a decay time of 200~ps. The dependencies of the recombination times $\tau_{\rm R1}$ and $\tau_{\rm R2}$ on $P$ are shown in Figure~\ref{figS2}c. We assign the longer recombination time $\tau_{\rm R2}$ to the recombination of spatially separated electrons and holes. As their concentrations increase at higher excitation densities, the separation between them reduces so that their recombination becomes faster. We suggest that the increase of $\tau_{\rm R1}$ time with growing $P$ evidences an increasing contribution of separated electrons and holes, compared to the exciton recombination. At the largest excitation density the PL dynamics is mostly related to electron-hole recombination.  



Figure~\ref{figS2}b shows the dynamics of optical orientation ($P_{\rm oo}(t)$) measured at $P=10$~mW/cm$^2$ and 600~mW/cm$^2$. The dependence at $P=10$~mW/cm$^2$ (blue dots) is the same as shown in the main text in Figure~2b, but here it extends over a longer temporal range up to 1.8~ns, highlighting the almost negligible decay of polarization across the whole time range. One can see, that at the initial time moment right after excitation $P_{\rm oo}(t=0)=0.85$ and then decreases and saturates at a level of about 0.60. The transition lasts about 55~ps, i.e. the same time that we have measured for the exciton lifetime in the recombination dynamics $\tau_{\rm X}=55$~ps. Therefore, the dynamics of $P_{\rm oo}(t)$ reflect the fact that the excitons with the very high optical orientation degree of 0.85 recombine and at longer time the orientational degree is given by the smaller values typical for resident carriers. Therefore, we observe here a rather unusual situation where the dynamics of $P_{\rm oo}(t)$ is determined not by the spin relaxation time, but by the recombination time. 

At the maximal $P=600$~mW/cm$^2$ used in experiment, the initial optical orientation decreases to $P_{\rm oo}(t=0)=0.50$. It drops over a time range of 150~ps after which it reaches the saturation level of about 0.15, see the black dots in Figure~\ref{figS2}b. The power dependences of $P_{\rm oo}(t=0)$ and $P_{\rm oo}(t=1.5~{\rm ns})$ are shown in Figure~\ref{figS2}d. One can see that the initial degree decreases steadily with increasing power density from 0.85 down to 0. while the degree at longer delays, which is contributed by resident carriers, shows the strongest decrease from 0.60 down to 0.20 up to powers of $100$~mW/cm$^2$. The decrease of the saturation level is related to an increase of the (short) exciton recombination time $\tau_{\rm R1}$ and is in line with Eq.~(2) of the main text. Indeed, at low pumping, the factor $\nu = \tau_s/(\tau_{\rm X}+\tau_s) \approx 0.8$ (from the experimental data), in consistency with $\tau_{\rm X} \equiv \tau_{\rm R1} =55$~ps and $\tau_s=220$~ps as discussed in the main text. It can describe the reduction of $P_{\rm oo}$ from 0.85 to 0.60, while the factor $\nu \approx 0.5$ corresponds to $\tau_{\rm X} \equiv \tau_{\rm R1} = 200$~ps  at high pump powers. Additional mechanisms such as a reduction of the exciton spin relaxation time at elevated densities cannot be excluded as well.


\section{Theory of exciton and carrier spin precession in magnetic field}

The cubic perovskites represent a special class of systems where the bottom conduction and topmost valence bands are `simple', two-fold spin degenerate. This results in quite specific exciton spin dynamics in magnetic field, controlled by the interplay of the exchange interaction and the Zeeman splittings of individual charge carriers. The basics of exciton spin physics in such a band structure in the presence of a transverse magnetic field were described in~\cite{excitons:RS}, see also~\cite{Odenthal:2017aa}. Here we present the theory with emphasis on the temporal dynamics of the exciton spin polarization. We would like to point out that the spin properties of bulk perovskite with cubic symmetry are quite different as compared with conventional III-V or II-VI semiconductors. The simple band structure of perovskites results in entirely new spin properties of the excitons.
In bulk III-V and II-VI semiconductors, the spin dynamics of excitons are difficult to study because the complex structure of the valence band and polariton effects play an important role.\cite{excitons:RS} In quantum wells based on conventional semiconductors, the exchange interaction separates the bright exciton states ($\pm 1$) from the dark exciton states ($\pm 2$) and strongly suppresses the in-plane magnetic field effect. In contrast to conventional III-V and II-VI semiconductors, in halide perovskites the bands are ``simple'' and the magnetic field efficiently couples the states of the bright exciton triplet.\cite{Tamarat2019,Tamarat2023}



\subsection{Temporal dynamics}

We recall that both the top valence and the bottom conduction bands in the lead halide perovskite semiconductors are two fold spin degenerate and can be conveniently described by spin-$1/2$ operators. The Hamiltonian of an electron-hole pair in perovskites with cubic symmetry in an external magnetic field $\bf{B}$ can be written as:
\begin{equation}
\label{XH}
\hat{H} = \mu_\text{B}g_\text{e}\mathbf{s}_{\rm e} {\bf{B}} + \mu_\text{B}g_\text{h}\mathbf{s}_{\rm h} {\bf{B}}+ \Delta_\text{X} \hat{\mathbf{s}}_{\rm e} \cdot \hat{\mathbf{s}}_{\rm h}.
\end{equation}
Here $\mu_\text{B}$ is the Bohr magneton, $\bf{B}$ is the external magnetic field, $\Delta_\text{X}$ is the electron-hole exchange splitting, $\mathbf s_\text{e}$ and $\mathbf s_\text{h}$ are the electron and hole spin-$1/2$ operators. Note that the exchange interaction, in agreement with the cubic symmetry, can be recast as
\[
\hat{\mathbf{s}}_{\rm e} \cdot \hat{\mathbf{s}}_{\rm h} = \frac{1}{2} \hat{\mathbf J}^2 - \frac{3}{4},
\]
where $\hat{\mathbf J} = \hat{\mathbf{s}}_\text{e} + \hat{\mathbf{s}}_\text{h}$ is the total spin operator of the electron hole pair and $\hat{\mathbf J}^2 = J(J+1)$. The $J$ takes one of the two values $0$ or $1$.  In what follows, we choose the eigenstates of the exchange interaction Hamiltonian in the form of $\lvert J,J_z\rangle$ as:
\begin{subequations}
\label{basic:4}
\begin{align}
\label{eq:fi1}
&\phi_1 = \lvert 1,+1\rangle = \lvert \uparrow, \Uparrow \rangle,\\
&\phi_2 = \lvert 1,0\rangle = \frac{1}{\sqrt{2}}(\lvert \uparrow, \Downarrow \rangle + \lvert \downarrow, \Uparrow \rangle),\\
&\phi_3 = \lvert 1,-1\rangle = \lvert \downarrow, \Downarrow \rangle,\\
\label{eq:fi4}
&\phi_4 = \lvert 0,0\rangle = \frac{1}{\sqrt{2}}(\lvert \uparrow, \Downarrow \rangle - \lvert \downarrow, \Uparrow \rangle).
\end{align}
\end{subequations}
Here, the up and down arrows indicate the spin: $+1/2$ and $-1/2$, respectively. The symbols $\uparrow$ and $\Uparrow$ correspond to electron and hole spins, respectively. In what follows, it is convenient to take the $z$-axis along the $\mathbf k$-vector of light. At $ B=0$ the eigenstates are the spin singlet, $J=0$, with the energy $-3\Delta_{\rm X}/4$ and the spin triplet, $J=1$, with the energy $\Delta_{\rm X}/4$, the splitting between these states is $\Delta_{\rm X}$. 
The states with $J_z= \pm 1$ ($\phi_1$, $\phi_3$) are optically active in $\sigma^\pm$ circular polarizations. The exciton state $|1,0\rangle$ has the dipole moment along the $z$-axis (``longitudinal'' exciton) and dark in this geometry. The state $|0,0\rangle$ is dark (spin forbidden).

In the presence of a magnetic field, the states with the same component of the total spin along the magnetic field are mixed, and the states with different components of the total spin get split. Particularly, in the Voigt geometry where $\mathbf B = (B_\text{V}, 0, 0) \parallel x$, the Hamiltonian~\eqref{XH} takes the form
\begin{equation}
\hat{H} = \frac{1}{2\sqrt{2}}
\begin{pmatrix}
2\sqrt{2}\Delta_\text{X} & \mu_\text{B}g_\text{V,X}B_\text{V} & 0 & -\mu_\text{B}g_\text{V,DX}B_\text{V}\\
 \mu_\text{B}g_\text{V,X}B_\text{V} & 2\sqrt{2}\Delta_\text{X} & \mu_\text{B}g_\text{V,X}B_\text{V} & 0\\ 
 0 & \mu_\text{B}g_\text{V,X}B_\text{V} & 2\sqrt{2}\Delta_\text{X} & \mu_\text{B}g_\text{V,DX}B_\text{V}\\ 
 -\mu_\text{B}g_\text{V,DX}B_\text{V} & 0 & \mu_\text{B}g_\text{V,DX}B_\text{V} & 0\\
\end{pmatrix},
\end{equation}
where $g_\text{V,X} = g_\text{V,e} + g_\text{V,h}$ is the so-called bright exciton $g$-factor and $g_\text{V,DX} = g_\text{V,e} - g_\text{V,h}$  is the dark exciton $g$-factor. Note that while in the Voigt geometry all excitonic states are optically active, in general, see the eigenfunctions of the Hamiltonian below, we use the notations of the bright and dark exciton $g$-factors in analogy with the Faraday geometry. The Larmor frequency of the bright exciton is defined by $\omega_\text{L,X} =  \mu_\text{B}g_\text{V,X}B_\text{V}/\hbar$.

The energies of the exciton states in the presence of the field read
\begin{subequations}
\label{energ:V}
\begin{gather}
\label{eq:Ej1}
E_\text{I} = \frac{1}{2}\left(\Delta_\text{X} - \sqrt{\Delta_\text{X}^2 + (\mu_\text{B}g_\text{V,DX}B_\text{V})^2}\right),\\
E_\text{II} = \frac{1}{2}\left(\Delta_\text{X} + \sqrt{\Delta_\text{X}^2 + (\mu_\text{B}g_\text{V,DX}B_\text{V})^2}\right),\\
\label{eq:Ej2}
E_\text{III} = \Delta_\text{X} - \frac{1}{2}\mu_\text{B}g_\text{V,X}B_\text{V},\\
E_\text{IV} = \Delta_\text{X} + \frac{1}{2}\mu_\text{B}g_\text{V,X}B_\text{V}.
\end{gather}
\end{subequations}

The eigenfunctions can be conveniently expressed as superpositions of the basic states~\eqref{basic:4} as
\begin{equation}
\label{basic:b}
|\mathrm i\rangle = \sum_j a_{\mathrm i,j} \phi_j,
\end{equation}
where the Roman index $\mathrm i=$ I, II, III or IV denotes the eigenstates in the magnetic field, and the arabic subscript $j=1,2,3,$ and 4 denotes the basic functions in \eqref{basic:4}. The decomposition coefficients $a_{\mathrm i,j}$ can be recast in column-vector form: 
\begin{equation}
\label{eq:a11}
a_{\mathrm I, j} = 
\begin{pmatrix}
\frac{1}{2}\sqrt{1-\Delta_\text{X}/C}\\
0\\ 
-\frac{1}{2}\sqrt{1-\Delta_\text{X}/C}\\ 
\frac{\mu_\text{B}g_\text{V,DX}B_\text{V}}{\sqrt{2C(C-\Delta_\text{X})}}\\
\end{pmatrix}, \quad 
a_{\mathrm{II}, j} = 
\begin{pmatrix}
-\frac{1}{2}\sqrt{1+\Delta_\text{X}/C}\\
0\\ 
\frac{1}{2}\sqrt{1+\Delta_\text{X}/C}\\ 
\frac{\mu_\text{B}g_\text{V,DX}B_\text{V}}{\sqrt{2C(C+\Delta_\text{X})}}\\
\end{pmatrix}, \quad 
a_{\mathrm{III}, j} = 
\begin{pmatrix}
\frac{1}{2}\\
-\frac{\sqrt{2}}{2}\\ 
\frac{1}{2}\\ 
0\\
\end{pmatrix}, \quad 
a_{\mathrm{IV}, j} = 
\begin{pmatrix}
\frac{1}{2}\\
\frac{\sqrt{2}}{2}\\ 
\frac{1}{2}\\ 
0\\
\end{pmatrix},
\end{equation}
with $C = \sqrt{\Delta_\text{X}^2 + (\mu_\text{B}g_\text{V,DX}B_\text{V})^2}$. 

The dependences of the exciton level energies on $B_\text{V}$ are shown in Figure~\ref{fig5}a for the case of zero exchange interaction $\Delta_\text{X} = 0$\,meV. In that case, a linear-in-field Zeeman splitting is seen for all four states. In the presence of a significant exchange splitting, $\Delta_\text{X} = 0.42$\,meV, the pair of states III and IV show a linear-in-$B_{\rm V}$ Zeeman splitting. The pair of states I and II demonstratea a linear-in-field splitting only at high fields with an offset given by $\Delta_{\rm X}$ for $B_{\rm V} \to 0$, as shown in Figure~\ref{fig5}d.

\begin{figure}[hbt]
\begin{center}
\includegraphics[width = 18cm]{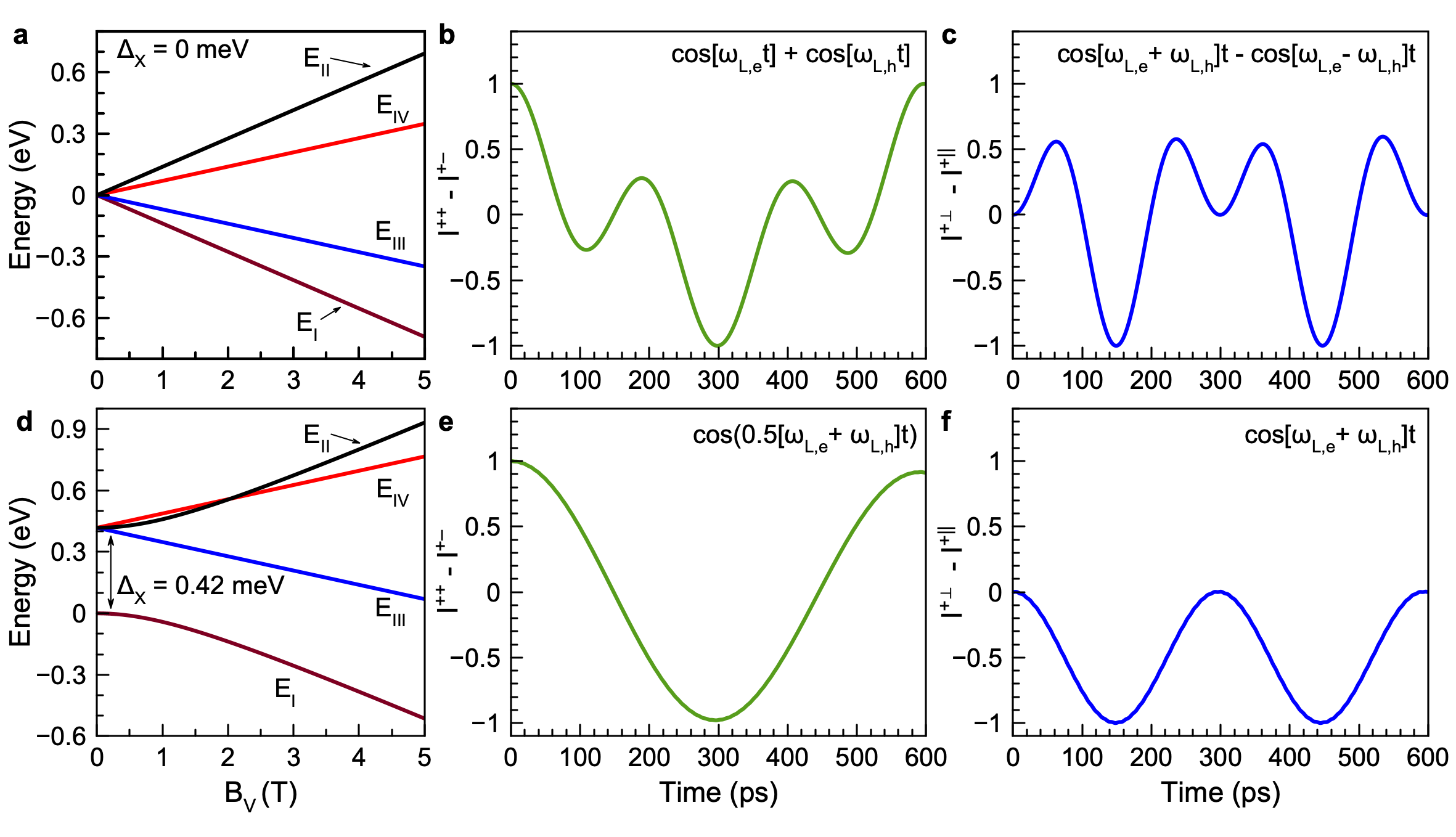}
\caption{\label{fig5} 
{Panels (a) and (d) show the energy levels of the excitons as a function of the magnetic field $B_\text{V}$, calculated for $\Delta_\text{X} = 0$\,meV (a) and $\Delta_\text{X} = 0.42$\,meV (d). Panels (b) and (e) present the dynamics of the signal $I^+ - I^-$ calculated for $\Delta_\text{X} = 0$\,meV (b) and $\Delta_\text{X} = 0.42$\,meV (e). Panels (c) and (f) show the dynamics of $I^{\text{V}} - I^{\text{H}}$ calculated for $\Delta_\text{X} = 0$\,meV (c) and $\Delta_\text{X} = 0.42$\,meV (f). Calculations are done for $g_\text{e,V} = +3.48$, $g_\text{h,V} = -1.15$, $B_\text{V} = 0.1$~T.}}
\end{center}
\end{figure}
 
In order to calculate the time- and polarization-resolved intensity of the photoluminescence following polarized excitation by a short laser pulse, we resort to a coherent model and disregard both the finite lifetime of excitons and the spin relaxation processes. Accordingly, we write the wavefunction of the system $\Psi(t)$ as a superposition of the eigenstates $|\mathrm i\rangle$ in Eq.~\eqref{basic:b}:
\begin{equation}
\Psi(t) = \sum_{\mathrm i =\text{I}}^{\text{IV}}C_{\mathrm i} |\mathrm i\rangle \exp(-i\omega_{\mathrm i} t), \quad \omega_{\mathrm i} = E_{\mathrm i}/\hbar.
\end{equation} 
The decomposition coefficients $C_{\mathrm i}$ are determined by the initial conditions: excitation of the system by a circularly polarized light renders excitons in the $\phi_1$ or $\phi_3$ states, depending on the photon helicity, and the initial state $(\phi_1 -\phi_3)/\sqrt{2}$ corresponds to horizontal polarization along $B_\text{V}$, marked by $\parallel$.  In the same fashion, $(\phi_1 + \phi_3)/\sqrt{2}$ describes the vertical, marked by $\perp$ , polarized excitation. Similarly, the intensity of the emission in the given polarization is determined by the absolute value squared of the $\Psi(t)$ projection onto the correspondingly polarized state. In our experiments, the intensities in the $\sigma^+$, $\sigma^-$, $\parallel$ and $\perp$ polarizations were measured after $\sigma^+$ polarized excitation. They read
\begin{subequations}
\begin{align}
&I^{++} = \left|\sum_{\mathrm i}|a_{{\mathrm i},1}|^2 e^{-i\omega_{\mathrm i}t}\right|^2,\\
&I^{+-} = \left|\sum_{\mathrm i} a_{{\mathrm i},1}a_{{\mathrm i},3} e^{-i\omega_{\mathrm i}t}\right|^2,\\
&I^{+\parallel} = \left|\sum_{\mathrm i} (a_{{\mathrm i},1}^2 + a_{{\mathrm i},1}a_{{\mathrm i},3}) e^{-i\omega_{\mathrm i}t}\right|^2,\\
&I^{+\perp} = \left|\sum_{\mathrm i} (a_{{\mathrm i},1}^2 - a_{{\mathrm i},1}a_{{\mathrm i},3}) e^{-i\omega_{\mathrm i}t}\right|^2.
\end{align}
\end{subequations}
Overall, the calculated intensities in Figure~\ref{fig5} show, in general, a complex oscillatory dependence on time related to the superposition of quantum beats with the frequencies following from Eqs.~\eqref{energ:V}. For the beats in circular polarization, simplified analytical expressions can be derived in the limit of negligible exchange, $\Delta_{\rm X} \to 0$, where 
\begin{gather}
\label{DCP_QB}
I^{++}  - I^{+-}  \propto \cos(\omega_\text{L,e}t) + \cos(\omega_\text{L,h}t),
\end{gather}
here $\omega_{\rm L,e(h)} = g_{\rm e(h)} \mu_\text{B} B_\text{V}/\hbar$ are the electron and hole Larmor precession frequencies. Naturally, in the case of $\Delta_{\rm X} \to 0$, excitonic effects are absent and the spin precessions observed correspond to those of the electron and hole spins with corresponding Larmor frequencies. in agreement with the numerical result in Figure~\ref{fig5}b. Such situation also holds for the ``long-living'' TRPL signal caused by unbound electrons and holes, see Figure~3a of the main text. If the exchange splitting is large as compared to the Zeeman splitting, see $\Delta_\text{X} > \hbar\omega_\text{L,X}$, then we can disregard the admixture of the singlet exciton and analyze the spin beats of the triplet of states described by the Zeeman Hamiltonian
\begin{equation}
\label{triplet}
\mathcal H_3 = \mu_B \frac{g_{\rm V,X} }{2} B_{\rm V} \hat L_x,
\end{equation} 
where $\hat L_x$ is the matrix of the $x$-component of the angular momentum $1$. Thus, the exciton pseudospin precesses with the frequency $\omega_{\rm L,X}/2 =  \mu_B g_{\rm V,X} B_{\rm V}/(2\hbar)$ and
\begin{gather}
\label{DCP_QB:strong}
I^{++}  - I^{+-}  \propto \cos(\omega_\text{L,X}t/2),
\end{gather}
in agreement with the beats calculated numerically in Figure~\ref{fig5}e for weak magnetic fields. In case of a strong exchange splitting the  $I^{++}  - I^{+-}$ signal oscillates with half the Larmor frequency of the Zeeman splitting of the bright exciton $\omega_\text{L,X}/2$. 

Similarly, in the limit of weak exchange interaction the beats in linear polarization can be described by the simple expression
\begin{equation}
I^{+\perp} \propto 1 + \cos[(\omega_\text{L,e} + \omega_\text{L,h})t],\quad
I^{+\parallel}  \propto 1 + \cos[(\omega_\text{L,e} - \omega_\text{L,h})t].
\end{equation}
The vertically linearly polarized component of the PL oscillates with the Larmor frequency of the bright exciton and the horizontally linearly polarized component with the dark exciton Larmor frequency. So, the signal $I^{\perp} - I^{\parallel}$ contains two oscillating contributions at the $\omega_\text{L,e} + \omega_\text{L,h}$ and $\omega_\text{L,e} - \omega_\text{L,h}$ frequencies, as shown in the Figure~\ref{fig5}c. In the case of strong exchange interaction the analysis within the Hamiltonian~\eqref{triplet} gives rise to the beats
\begin{equation}
\label{beats:lin:strong}
I^{+\perp}  \propto 0.5(1 + \cos[\omega_{\rm L,X}t]),\quad
I^{+{\parallel} } \propto 1,
\end{equation}
in agreement with the numerical calculations. In the experiments, exciton beats are seen in the linear polarization, see Figure~4c, and are reasonably welldescribed by Eq.~\eqref{beats:lin:strong}.

We stress that the expressions for the temporal dependence of the intensities are derived within a fully coherent approach that disregards the spin relaxation and population relaxtion of excitons. Phenomenologically, these can be included by the exponential factors $\exp{(-t/\tau_{\rm X})}$ to account for the exciton recombination and $\exp{(-t/\tau_{s})}$ to account for the spin relaxation.

The model described above can be applied to the analysis of both the fast dynamics of excitons with large exchange splitting and charge carriers with negligible exchange interaction. The exciton dynamics occur mainly on the exciton radiative lifetime scale under the influence of the exchange interaction and Zeeman splittings. The dynamics of long-living electrons and holes is defined by the Zeeman splittings of individual charge carriers.

\subsection{Electron and hole spin correlations in the linear polarization of emission}\label{sec:correlations:end}

It is instructive to relate the linear polarization of emission with the correlators of electron and hole spins. To that end, we focus on the triplet $J=1$ which consists, as described above, of two states active in $\sigma^\pm$ polarization for light propagating along the $z$-axis and the ``longitudinal'' state. This triplet is described by the operators $\hat{L}_\alpha$ of the angular momentum-$1$, where $\alpha=x,y,z$. One can directly check that the degree of \emph{circular} polarization of single exciton emission is given by the quantum mechanical average
\begin{subequations}
\label{angular:Stokes}
\begin{equation}
\label{angular:Stokes:circ}
P_{\rm oo} = \langle \hat{L}_z\rangle,
\end{equation}
while the degrees of \emph{linear polarization} in the sets of axes $(xy)$ and $(x'y')$ rotated by $45^\circ$ relative to each other are given by
\begin{align}
P_{\rm lin} &= \langle \hat{L}_x^2-\hat{L}_y^2\rangle,\label{angular:Stokes:lin} \\
P_{\rm lin}' &= \langle \hat{L}_x \hat{L}_y +  \hat{L}_y \hat{L}_x  \rangle.\label{angular:Stokes:lin1} 
\end{align}
\end{subequations}
Making use of the fact that the total exciton spin $\mathbf J = \hat{\mathbf s}_{\rm e}+ \hat{\mathbf s}_{\rm h}$ and that the state with $J=0$ is optically inactive, one can readily check that 
\begin{subequations}
\label{angular:Stokes:S}
\begin{align}
P_{\rm oo} &= \langle \hat{s}^{\rm e}_z+\hat{s}^{\rm h}_z\rangle,\label{angular:Stokes:circ:S}\\
P_{\rm lin} &= 2\langle \hat{s}_{x}^{\rm e} \hat{s}_{x}^{\rm h} - \hat{s}_{y}^{\rm e} \hat{s}_{y}^{\rm h}\rangle,\label{angular:Stokes:lin} \\
P_{\rm lin}' &= 2\langle \hat{s}_{x}^{\rm e} \hat{s}_{y}^{\rm h} + \hat{s}_{y}^{\rm e} \hat{s}_{x}^{\rm h} \rangle.\label{angular:Stokes:lin1} 
\end{align}
\end{subequations}
Here, to avoid confusion, we use superscipts to denote the type of particle $\rm e$ or $\rm h$ in the exciton and subscripts to denote the Cartesian components. 

Equations~\eqref{angular:Stokes:S} clearly demonstrate that to obtain circular polarization of emission, the polarization of electron or hole is sufficient. At the same time, linear polarization requires the correlation of the electron and hole spins. Note that the approach with the exciton spin operator $\hat{\mathbf L}$ allows one to derive (in the limiting case of a sufficiently high exchange interaction) the set of coupled equations for the exciton polarization in agreement with Ref.~\citenum{excitons:RS}.

Here we mention two possible scenarios of PL generation and exciton recombination (see also concluding paragraphs of Sec.~\ref{sec:oo_calcul}): excitons that emit light can either be formed (i) from (partially) polarized electrons and holes that retain some degree of polarization during energy relaxation or (ii)  from unpolarized resident carriers and polarized photocarriers. The comparison of the linear and circular polarization degrees in Figures~4a and 4c in the main text shows that the number of excitons formed by mechanism (i) is smaller than the number of those from (ii). The process (i) probably corresponds to geminate exciton formation where the electron-hole pair formed by photon absorption binds into an exciton. Further measurements including the degree of linear polarization for linearly polarized excitation (exciton alignment) and comparison of the total emission intensity in case of circular and linear excitation [cf. Ref.~\citenum{PhysRevB.50.11624}] may further elucidate the processes, geminate or bimolecular, of exciton formation in perovskite crystals.

\subsection{Hanle effect}

The spin dynamics of excitons and resident carriers occur on different time scales. Hence, to analyze the exciton Hanle effect extracted from the dynamics of the degree of optical orientation, it is necessary to develop a suitable theoretical model. In the experiment, we measure $I^{++}$ and $I^{+-}$, the intensities of the $\sigma^+$ and $\sigma^-$ polarized components of the PL, excited by $\sigma^+$ polarized light. To calculate the degree of optical orientation, we used the ratio $$P_{\rm oo}= \frac{I^{++} - I^{+-}}{I^{++} + I^{+-}}.$$ The difference of intensities $I^{++} - I^{+-}$ is determined by the spin density of excitons or charge carriers $S_z$, and the total intensity $I^{++} + I^{+-}$ corresponds to the population $N(t)$.

The Hanle effect results from depolarization of the photoluminescence in a transverse magnetic field applied in the Voigt geometry ($\mathbf{B}_\text{V} \perp \mathbf{k}$). Here, we briefly discuss the contributions of (i) resident electrons and holes and (ii) excitons to the Hanle effect.

\begin{figure}[htb]
\begin{center}
\includegraphics[width = 6cm]{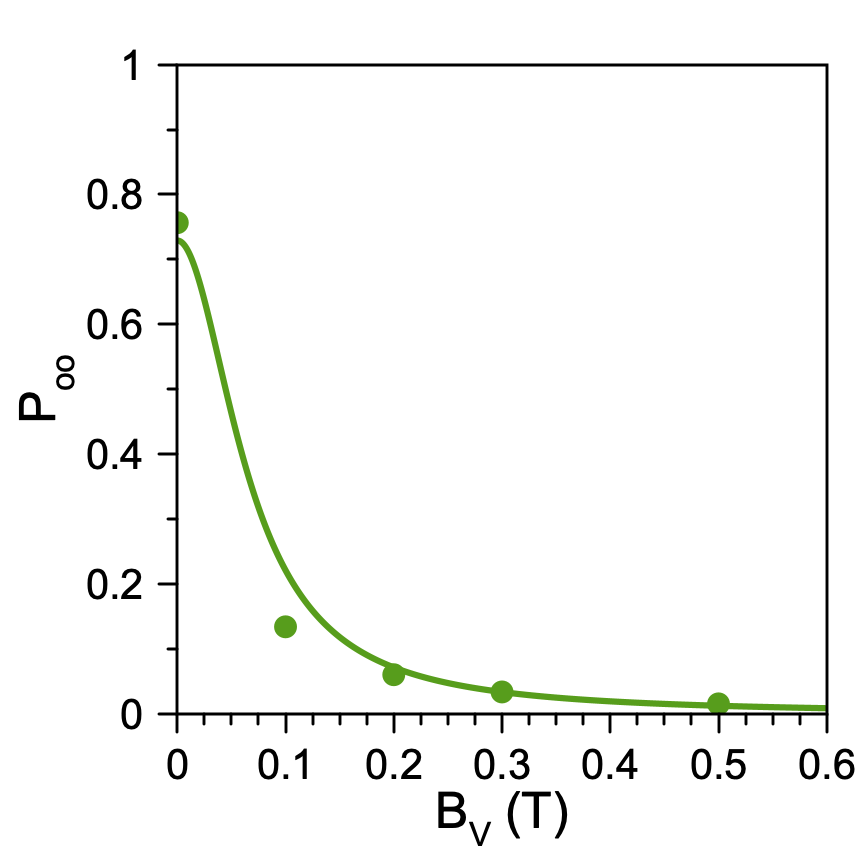}
\caption{\label{Hanle} 
Hanle effect of exciton depolarization in a transverse magnetic field. $P_\text{oo}$ dependence on $B_\text{V}$ is shown by the dots measured for $\sigma^+$ excitation at $E_\text{exc} = 1.675$\,eV, $E_\text{det} = 1.508$\,eV, $P = 30$\,mW/cm$^2$ and $T = 1.6$\,K. The fitting by Eq.~\eqref{eq:H8} gives the $T_\text{s} = 75$\,ps and $\tau_\text{X} = 85$\,ps.} 
\end{center}
\end{figure}

The contribution of long-living charge carriers to the Hanle effect can be described by the standard expressions, see, e.g., Refs.~\citenum{opt_or_book,glazov2018electron}. For long spin relaxation times exceeding nanoseconds, the electron and hole Hanle curves are extremely narrow: 4~mT for electrons and 18~mT for holes. The free carrier contribution to $P_{\rm oo}(B_{\rm V})$ is not observed in the range of fields up to a $0.5$~T used in the experiment, see Fig. 2f of the main text. 

To isolate the exciton contribution, we integrate the measured intensities $I^{++}$ and $I^{+-}$ over time up to $t_1 = 600$~ps which exceeds the exciton lifetime, but is much shorter than the carrier spin relaxation time. In this way, we determine the values $P_{\rm oo}(B_{\rm V})$ plotted in Figure~\ref{Hanle}a. To analyze the experimental data, we assume that for the relevant magnetic fields ($B_{\rm V} \lesssim 100$~mT) the exchange interaction in the exciton is dominating in this regime. Thus, a simple calculation after Eq.~\eqref{DCP_QB:strong} shows that 
\begin{gather}
\label{eq:H8}
P_\text{oo}(B_\text{V}) = \frac{\tau_\text{s}}{\tau_\text{X}+\tau_\text{s}}\cdot \frac{P_\text{oo}(t=0)}{1 + (\mu_\text{B}g_\text{V,X}B_\text{V}T_\text{s}/2\hbar)^2} ,
\end{gather}
where $T_s = \tau_{\rm X}\tau_{\rm s}/(\tau_{\rm X} + \tau_{\rm s})$ is the exciton spin lifetime with $\tau_{\rm s}$ being the exciton spin relaxation time, and $P_\text{oo}(0)$ is the optical orientation degree of excitons in zero field. This equation describes the Hanle curve for continuous wave excitation or case of $T_\text{s}$ being much shorter than the integration time of the TRPL dynamics. Naturally, the steady-state $P_\text{oo}$ at $B_\text{V} = 0$~T is given by $P_\text{oo}(t=0)\tau_\text{s}/(\tau_\text{s} + \tau_\text{X})$.

Fitting the $P_\text{oo}$ dependence on $B_\text{V}$ with Eq.~\eqref{eq:H8} gives the half-width at half-maximum $\delta B_{\rm{X}} = 2\hbar/(\mu_\text{B}g_\text{V,X}T_\text{s}) = 65$~mT and $T_\text{s} = 150$~ps, calculated with $g_\text{V,X} = 2.3$.  Note that the estimation of $T_\text{s}$ from the Hanle half-width is a rough approximation, since the use of this expression means that the integration time of the experimental signals is much longer than the spin lifetime. 
The width of the Hanle curve of $\delta B_\text{X} = 65$~mT corresponds to $T_\text{s} = 75$~ps. Using the exciton lifetime $\tau_\text{X} = 85$~ps measured at $P = 30$~mW/cm$^2$, we obtain $\tau_\text{s} = 300$~ps.  

Note that the very large value of $P_{\rm oo}=0.6$, which remains constant in the temporal range from 100 to 500~ps (in fact, we have measured this value up to 2~ns), evidences remarkably long spin relaxation times of electrons and holes (at least for one of this carrier, but most probably for both). Using as an estimate for the electron-hole recombination time the lower limit $\tau_\text{R2}=840$~ps, we obtain $\tau_\text{s,car} \ge 1.5\tau_\text{R2}=1260$~ps.

\section{Calculations}

\subsection{Band structure calculations}

For the calculations, we use the empirical tight-binding (ETB) approach in its nearest-neighbor $sp^3d^5s^*$ version.\cite{Jancu1998} It allows us to reproduce the dispersion calculated in state-of-art density functional theory (DFT) with meV precision for the few bands near the band gap.\cite{Nestoklon2021} For the band structure calculations we relied on the modified Becke-Johnson exchange–correlation potential\cite{Tran2009} in Jishi parametrization\cite{Jishi2014} as implemented in the WIEN2k package.\cite{WIEN2k}

\subsection{Absorption}
Let us first remind how the light absorption is calculated as the calculation of the depolarization has a similar structure. The light absorption up to a constant equals to the imaginary part of the dielectric permittivity. The imaginary part of the polarization-averaged dielectric function is given by \cite{Graf1995}
\begin{equation}\label{eq:epsiloni}
  {\rm Im}\left\{ \varepsilon(E) \right\} = \frac{4\pi\alpha\hbar c}{V} \sum_{i,a,{\bf k}} \frac{\frac13\left| \braket{i,{\bf k}}{\bf v}{a,{\bf k}} \right|^2 }{E^2}
  \delta\left[ E-E_a({\bf k})+E_i({\bf k}) \right]\,,
\end{equation}
where $\alpha$ is the fine structure constant, $c$ is the light velocity, $V$ is the unit cell volume, $\braket{i,{\bf k}}{\bf v}{a,{\bf k}}$ is the matrix element of velocity calculated with the states $\ket{i,{\bf k}}$ in the valence and $\ket{a,{\bf k}}$ in the conduction bands, $E_i({\bf k})$ and $E_a({\bf k})$ are the energies of the conduction and valence band states with wave vector ${\bf k}$. In numerical calculations, the $\delta$-function is approximated with a Gaussian with fixed width. We compute $\braket{i,{\bf k}}{\bf v}{a,{\bf k}}$ in the tight-binding approach using the parameters given in Table~\ref{tbl:TB_par}. We implement the parameters from Ref.~\citenum{Nestoklon2021}, fitted to reproduce the band structure obtained in DFT calculations, but change the diagnoal energy of the $p$-orbital of Pb ($E_{pc}$) to fit exactly the experimental band gap. We carefully checked that the polarization as a function of energy is not sensitive to such small changes of the band gap. 

\begin{table} [htb]
\caption{Tight-binding parameters used in the calculations. In addition to the parameters presented in the table, $s_cs^*_a\sigma$, $s^*_ad_c\sigma$, $p_cd_a\sigma$, $p_ad_c\pi$ and the parameters involving the $s_a$ and $s^*_c$ orbitals are assumed to be zero.}
\label{tbl:TB_par}
\begin{tabular}{lrr}
 \hline
 \hline
 &  Ref.~\citenum{Nestoklon2021} & expt. corrected \\
 \hline
$         E_{sc}$  & $   -5.7767$ &  $   -5.7767$ \\
$       E_{s^*a}$  & $   19.6780$ &  $   19.6594$ \\
$         E_{pa}$  & $   -2.3350$ &  $   -2.3350$ \\
$         E_{pc}$  & $    4.4825$ &  $    4.6587$ \\
$         E_{da}$  & $   10.8491$ &  $   10.8631$ \\
$         E_{dc}$  & $   13.9357$ &  $   13.9357$ \\
$   s_cp_a\sigma$  & $    1.0421$ &  $    1.0421$ \\
$ s^*_ap_c\sigma$  & $    2.7092$ &  $    2.8971$ \\
$   s_cd_a\sigma$  & $    0.3749$ &  $    0.3751$ \\
$       pp\sigma$  & $   -1.8838$ &  $   -1.8687$ \\
$          pp\pi$  & $    0.1955$ &  $    0.0929$ \\
$   p_ad_c\sigma$  & $    1.0341$ &  $    1.0342$ \\
$      p_cd_a\pi$  & $   -0.7960$ &  $   -0.2733$ \\
$       dd\sigma$  & $   -1.1231$ &  $   -1.1231$ \\
$          dd\pi$  & $    2.0000$ &  $    2.0000$ \\
$       dd\delta$  & $   -1.4000$ &  $   -1.4000$ \\
$     \Delta_a/3$  & $    0.3250$ &  $    0.3250$ \\
$     \Delta_c/3$  & $    0.4892$ &  $    0.4892$ \\
 \hline
 \hline
 \end{tabular}
 \end{table}

The calculated absorption is shown in Figure~5a of the main text. It is in a good agreement with other calculations (see Ref.~\citenum{BoyerRichard2016} or more recent Ref.~\citenum{Song2020}). Note that the exciton effects are not included here. They would add a sharp peak right below the gap energy at the $R$ point\cite{Singh2016} and increase the absorption near the $M$ point. It is well known~\cite{BoyerRichard2016} that the absorption increases significantly from the band gap at the $R$ point up to the $M$ point. Figure~5b shows the calculated band dispersion for bulk {FA}$_{0.9}$Cs$_{0.1}$PbI$_{2.8}$Br$_{0.2}$ along the $\Gamma \to R \to M \to X$ path.

\subsection{Calculation of optical orientation}\label{sec:oo_calcul}

We recall that in experiment the circular (or linear) polarization of the emission in the exciton range of the spectrum is measured after non-resonant excitation, typically, well above the free-particle band gap. As a result, free electrons and holes with significant excess kinetic energies are generated, they loose their energy and eventually form excitons (direct in real space or separated due to potential fluctuations, see~Ref.~\citenum{kirstein2021} and references therein), before they recombine leading to light emission. Thus, to describe the optical orientation one needs to take into account three basic steps:
\begin{enumerate}

\item The circularly polarized optical pulse with energy $E_{\rm exc}$ and wave vector ${\bf k}_{\rm exc}$ transfers an electron from the valence band state with wavevector $\mathbf k_i$ to the conduction band state with wavevector $\mathbf k_a$. satisfying the energy and momentum conservation laws: $E_a({\bf k}_a)-E_i({\bf k}_i)=E_{\rm exc}$; ${\bf k}_a-{\bf k}_i = {\bf k}_{\rm exc}$. The large speed of light allows neglecting the photon wave vector and the momentum conservation law reduces to ${\bf k}_a={\bf k}_i\equiv {\bf k}_0 $ (note that we work in the electron representation here). For a fixed energy of incident light, there is a set of possible ${\bf k}_0$ satisfying $E_a({\bf k}_0)-E_i({\bf k}_0)=E_{\rm exc}$, see Figure~5e of the main text for illustration. The spin state of the photoexcited carriers is defined by the polarization of incident light and is described by the density matrix of the carriers $\hat{\rho}^{c,v}({\bf k}_0)$, see below. 

\item  Electrons and holes loose their kinetic energies, first mostly due to emission of optical phonons, and near the $R$ point due to emission of acoustic phonons. Assuming a definite mechanism of energy relaxation, one may compute the final density matrix of carriers $\hat{\rho}^{c,v}({\bf k}_R)$ in the $R$ point. 

\item The electrons and holes recombine at the $R$ point and the polarization of the emitted light may be calculated from $\hat{\rho}^{c,v}({\bf k}_R)$.

\end{enumerate}

Let us now analyze the spin states of the photocarriers in more detail. Let us focus on the conduction electron, the general formalism for the holes is the same. The conduction band Bloch state $\ket{i,{\bf k}}$ can be decomposed in the two basic spinors $\uparrow$ and $\downarrow$, corresponding to the free electron spin $z$ component $s_z = +1/2$ and $-1/2$, respectively:
\begin{subequations}
\label{Bloch:k}
\begin{equation}
\label{Bloch:k:1}
\ket{i,{\bf k}} =  e^{\mathrm i \mathbf k\cdot \mathbf r} \left[U_{i\mathbf k}(\mathbf r) \uparrow + V_{i\mathbf k}(\mathbf r) \downarrow \right].
\end{equation}
Here the $U_{i\mathbf k}(\bm r)$, $V_{i\mathbf k}(\bm r)$ are the periodic Bloch amplitudes. We consider a centrosymmetric phase of the crystal, hence, the counterpart state with the same momentum and the same energy but opposite spin orientation is 
\begin{equation}
\label{Bloch:k:2}
\ket{\bar{i},{\bf k}} = \mathcal P \mathcal T \ket{i,{\bf k}}
  =  e^{\mathrm i \mathbf k\cdot \mathbf r} \left[U_{i-\mathbf k}^*(\mathbf r) \downarrow - V_{i,-\mathbf k}^*(\mathbf r) \uparrow \right],
\end{equation}
where $\mathcal P$ and $\mathcal T$ are the space and time reversal symmetry operators. Hereafter, these two conduction band states are denoted by the spin $s=\pm 1/2$ subscript. Similar relations hold for the valence band states and we denote corresponding states in the doublet by the subscript $m=\pm 1/2$.
\end{subequations}
 
The general equations~\eqref{Bloch:k} are valid for any $\mathbf k$ in the Brillouin zone. In the vicinity of the $R$ point, where the conduction band minimum and the valence band maximum are located, the Bloch functions take the form:
\begin{subequations}
\label{Rpoint:c}
\begin{align}
\psi^c_{+1/2,R}(\mathbf r) = -\frac{\mathcal Z(\mathbf r)}{\sqrt{3}}\uparrow - \frac{\mathcal X(\mathbf r) + \mathrm i\mathcal Y(\mathbf r)}{\sqrt{3}}\downarrow,\\
\psi^c_{-1/2,R}(\mathbf r) = \frac{\mathcal Z(\mathbf r)}{\sqrt{3}}\downarrow - \frac{\mathcal X(\mathbf r) - \mathrm i\mathcal Y(\mathbf r)}{\sqrt{3}}\uparrow,
\end{align}
\end{subequations}
for the conduction band and 
\begin{subequations}
\label{Rpoint:v}
\begin{align}
\psi^v_{+1/2,R}(\mathbf r) = \mathrm i \mathcal S(\mathbf r)\uparrow ,\\
\psi^v_{-1/2,R}(\mathbf r) = \mathrm i \mathcal S(\mathbf r)\downarrow,
\end{align}
\end{subequations}
for the valence band Bloch functions at the $R$ point. The notations $\mathcal S(\mathbf r)$, $\mathcal X(\mathbf r)$, $\mathcal Y(\mathbf r)$, and $\mathcal Z(\mathbf r)$ reflect the symmetry of the corresponding orbital states. The equations~\eqref{Rpoint:c} and \eqref{Rpoint:v} are consistent with the irreducible representations $R^{6,\mp}$ relevant for the conduction and valence band in the cubic perovskites.\cite{kirstein2021,kirstein2021nc}

The density matrix of charge carriers after photoexcitation is given by:\cite{opt_or_book}
\begin{subequations}\label{eq:rho}
\begin{align}
\label{rho_e}
\rho_{ss'}^{c}({\bf k}_0) = \frac{\sum_{m=\pm 1/2} [\mathbf v_{sm}({\bf k}_0) \cdot {\bf e}][\mathbf v_{s'm}({\bf k}_0) \cdot {\bf e}]^*}{\sum_{m=\pm 1/2,s=\pm 1/2} |\mathbf v_{sm}({\bf k}_0) \cdot {\bf e}|^2}\,,
\\
\label{rho_h}
\rho_{mm'}^{v}({\bf k}_0) = \frac{\sum_{s=\pm 1/2} [\mathbf v_{sm}({\bf k}_0) \cdot {\bf e}][\mathbf v_{s'm}({\bf k}_0) \cdot {\bf e}]^*}{\sum_{m=\pm 1/2,s=\pm 1/2} |\mathbf v_{sm}({\bf k}_0) \cdot {\bf e}|^2}
\end{align}\end{subequations} 
for the electrons and holes, respectively. Here $\mathbf e$ is the light polarization unit vector.

In Eqs.~\eqref{eq:rho}, the basis states of electron and hole are arbitrary. It is convenient to align them along the incident light polarization: 
For any light polarization $\gamma$ ($\gamma=+$ for right circular polarization, $\gamma=-$ for left circular polarization) given by the vector ${\bf e}^{\gamma}$ the matrix elements of the velocity projection onto the polarization vector ${\bf v}\cdot{\bf e}^{\gamma}$ are diagonal in the basis found by singular value decomposition (SVD). Its singular values are $v^{\gamma}_{\zeta}$ where $\zeta$ enumerates the ``pure" transitions. At the $R$ point of the Brillouin zone only one of them is non-zero, while both are generally non-zero for a ${\bf k}_{0}$ that does not correspond to any of high symmetry points. Both ${\bf v}\cdot{\bf e}^{+}$ and  ${\bf v}\cdot{\bf e}^{-}$ have the same common singular values and the states that diagonalize them ($v_0^+=v_1^-$ and $v_1^+=v_0^-$). In this basis, both density matrices are diagonal and equal. They reduce to
\begin{equation}\label{eq:rho_k0}
  \rho_{ss'}^{c+}({\bf k}_0) \equiv \rho_{ss'}^{v+}({\bf k}_0) = 
  \frac{|v_{s}^+({\bf k}_0)|^2}{|v_{0}^+({\bf k}_0)|^2+|v_{1}^+({\bf k}_0)|^2}\delta_{ss'}\,.
\end{equation}
In Eq.~\eqref{eq:rho_k0} and below, we consider only the right-circular polarization of excitation and omit the superscript $\gamma=+$. We stress that at the $R$ point the selection rules are such that circularly polarized light creates electrons and holes with pure states in the basis of Eqs.~\eqref{Rpoint:c}, \eqref{Rpoint:v}
\begin{equation}
\label{clear}
\sigma^+ \to (+1/2_e, +1/2_h), \quad \sigma^- \to (-1/2_e, -1/2_h).
\end{equation}
However, the band mixing at $\mathbf k \ne \mathbf k_R$ leads to a violation of the selection rules and an effective depolarization of charge carriers~\cite{opt_or_book,DP}.

The most complicated part of the calculation is the description of the evolution of the density matrix during energy relaxation of the carriers. Here we use two simplified approaches: (i) the ``effective phonon'' model, where we assume that there is a phonon mode which directly transfers electrons and holes from their initial state to the $R$ point of the Brillouin zone, and (ii) the ``effective emission'' model, where we assume that the polarization of the emission is determined by the selection rules in the excited states. In both cases we neglect excitonic effects, their role is briefly discussed after description of these models below.

\subsubsection{\texorpdfstring{``}{"}Effective phonon\texorpdfstring{''}{"} model}\label{subsec:effectivephonon}

In the effective phonon model we assume that there is an ``effective'' phonon with wavevector $\mathbf q = {\mathbf k_0} - \mathbf k_R$ and energy $\hbar\Omega_c(\mathbf q) = {E_c(\mathbf k_0)} -E_c (\mathbf k_R)$, which provides efficient scattering of the photoelectron from the initial state to the conduction band minimum. A similar phonon with $\hbar\Omega_v(\mathbf q) = E_v(\mathbf k_R) -{E_v (\mathbf k_0)}$ provides the hole scattering to the $R$ point from the photoexcited state. We assume that the interaction with such ``effective'' phonons is spin-independent. As a result, the loss of polarization is related to two effects:
\begin{enumerate}
\item Violation of the clean selection rules~\eqref{clear} at high wavevectors of the charge carriers, which are valid only at the R point. It results in spin depolarization of the photoexcited electrons. 
\item Mismatch of the spin orientation in the initial state $|i,\mathbf k_0\rangle$ and in the final state of a given carrier, Eqs.~\eqref{Rpoint:c} for the conduction band and Eqs.~\eqref{Rpoint:v} for the valence band in the phonon scattering (Elliott mechanism of spin relaxation, see Refs.~\citenum{Elliott54,liu2013spin} and Sec.~\ref{sec:spinrel} below).
\end{enumerate}

\begin{figure}[t]
\begin{center}
\includegraphics[width=\textwidth]{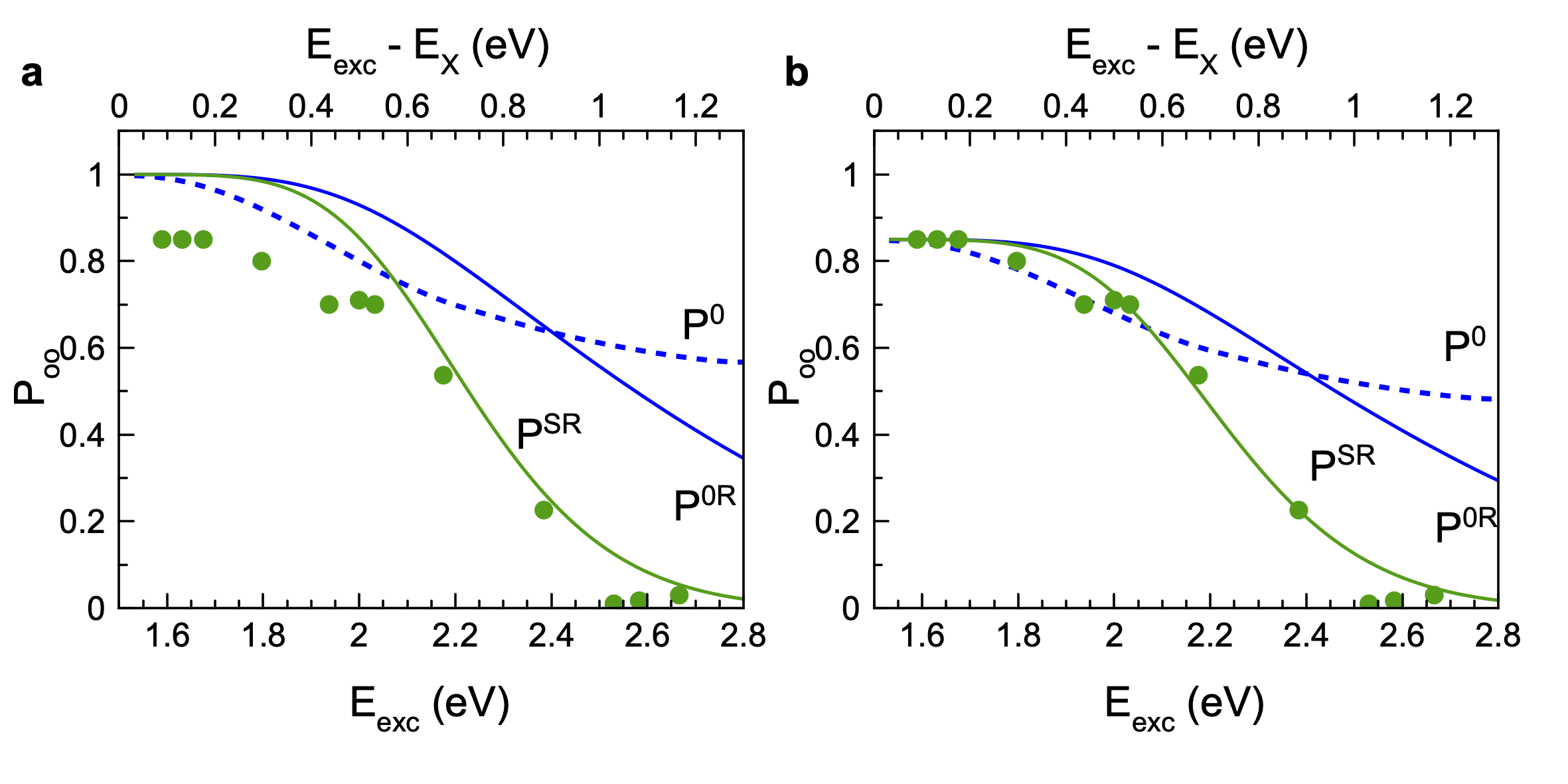}
\caption{Theoretically calculated $P^{\mathrm{0}}(E)$, see Eq.~\eqref{eq:P_E2} (dashed blue line), $P^{\rm 0R}(E)$, see Eq.~\eqref{eq:p_e_ph} (solid blue line) and $P^{\rm SR}(E)$, Eqs.~(\ref{eq:Ipm_general},\ref{eq:I_E},\ref{eq:rho_SR}) (solid green line). Points show the experimental values of the optical orientation degree as a function of the detuning energy. Panel (b) shows the same results, but the theory curves are multiplied by the depolarization factor $0.85$ to match the experimental value of $P_{\rm oo}=0.85$ for the small detuning energy of 0.1~eV.}\label{fig:PEs}
\end{center}
\end{figure}

Technically, the density matrix of the states at the $R$ point is found employing the transition matrices 
\begin{equation}\label{eq:overlaps_0R}
  V_{0R}^{\eta = c,v} \sim
  \begin{pmatrix}
    \braket{0_{\eta},{\bf k}_R}{e^{i({\bf k}_R-{\bf k}_{0}){\bf r}} }{0_{\eta},{\bf k}_{0}} &
    \braket{0_{\eta},{\bf k}_R}{e^{i({\bf k}_R-{\bf k}_{0}){\bf r}} }{1_{\eta},{\bf k}_{0}} \\
    \braket{1_{\eta},{\bf k}_R}{e^{i({\bf k}_R-{\bf k}_{0}){\bf r}} }{0_{\eta},{\bf k}_{0}} &
    \braket{1_{\eta},{\bf k}_R}{e^{i({\bf k}_R-{\bf k}_{0}){\bf r}} }{1_{\eta},{\bf k}_{0}}
  \end{pmatrix}
\end{equation}
which allow one to find the density matrix as 
\begin{equation}\label{eq:rho_R_one_phonon}
  \rho^{\eta}({\bf k}_R;{\bf k}_0) = V_{0R}^{\eta} \rho^{\eta}({\bf k}_0) \left( V_{0R}^{\eta}\right)^+
\end{equation}
and compute the intensities of light emitted in the left and right circular polarizations. 
Given the density matrices $\rho^{c,v}({\bf k})$ for carriers with arbitrary wave vector ${\bf k}$, written in the basis of circularly polarized states, the intensity of optical transitions in the right- and left-circular polarization may be calculated as:
\begin{subequations}\label{eq:Ipm_general}
\begin{align}
  I^+_{{\bf k}}({\bf k}_0) &= 
  \rho^c_{00}\rho^v_{00} |v_0({\bf k})|^2 + 
  \rho^c_{11}\rho^v_{11} |v_1({\bf k})|^2 + 
  \rho^c_{01}\rho^v_{10} v_0({\bf k}) v_1^*({\bf k}) + 
  \rho^c_{10}\rho^v_{01} v_1({\bf k}) v_0^*({\bf k})\,, 
\\
  I^-_{{\bf k}}({\bf k}_0) &= 
  \rho^c_{00}\rho^v_{00} |v_1({\bf k})|^2 + 
  \rho^c_{11}\rho^v_{11} |v_0({\bf k})|^2 + 
  \rho^c_{01}\rho^v_{10} v_1({\bf k}) v_0^*({\bf k}) + 
  \rho^c_{10}\rho^v_{01} v_0({\bf k}) v_1^*({\bf k})\,.
\end{align}
\end{subequations}

To calculate the polarization of output light, the intensity of light emitted in the two polarizations should be integrated over the wave vector (c.f. Eq.~\eqref{eq:epsiloni}): 
\begin{subequations}\label{eq:I_E}
\begin{align}\label{eq:P_E1}
  I^+_{{\bf k}}(E) = \int_{{\bf k}_0} I^+_{{\bf k}}({\bf k}_0) \delta\left( E-E_c\left( {\bf k}_0\right)+E_v\left( {\bf k}_0\right) \right)\,,
\\
  I^-_{{\bf k}}(E) = \int_{{\bf k}_0} I^-_{{\bf k}}({\bf k}_0) \delta\left( E-E_c\left( {\bf k}_0\right)+E_v\left( {\bf k}_0\right) \right)\,.
\end{align}
\end{subequations}
The polarization in this mode is then found from Eqs.~(\ref{eq:rho_R_one_phonon}-\ref{eq:I_E}) as
\begin{equation}\label{eq:p_e_ph}
  P^{\mathrm{0R}}(E) = \frac{I^+_{{\bf k}_R}(E) - I^-_{{\bf k}_R}(E)}{I^+_{{\bf k}_R}(E) + I^-_{{\bf k}_R}(E)}\,.
\end{equation}
The results of these calculations are shown in Figure~\ref{fig:PEs} by the solid blue line.
All matrix elements in Eqs.~(\ref{eq:rho_k0},\ref{eq:overlaps_0R}) as well as state energies are calculated in the empirical tight-binding approach. In calculations we take a $50\times50\times50$ $k$-mesh in $1/8$ of the (cubic) Brillouin Zone.

\subsubsection{\texorpdfstring{``}{"}Effective emission\texorpdfstring{''}{"} model}\label{subsec:effectiveemission}

In the ``effective emission'' model we calculate the polarization of emission from the photoexcited states, without an account of the energy relaxation (i.e., the optical orientation of hot excitons). Then, Eq.~\eqref{eq:Ipm_general} with the use of Eq.~\eqref{eq:rho_k0} reduces to 
\begin{subequations}\label{eq:Ipm_k0}
\begin{align}
  I^+_{{\bf k}_0}({\bf k}_0) &= 
  \frac{ |v_0|^6 + |v_1|^6  }{\left(|v_{0}^+({\bf k}_0)|^2+|v_{1}^+({\bf k}_0)|^2\right)^2}\,,
\\
  I^-_{{\bf k}_0}({\bf k}_0) &= 
  \frac{ |v_0|^4|v_1|^2 + |v_1|^4|v_0|^2  }{\left(|v_{0}^+({\bf k}_0)|^2+|v_{1}^+({\bf k}_0)|^2\right)^2}\,.
\end{align}
\end{subequations}
The upper boundary of the optical orientation degree may be calculated from Eqs.~(\ref{eq:I_E},\ref{eq:Ipm_k0}) as
\begin{equation}\label{eq:P_E2}
  P^{0}(E) = \frac{I^+_{{\bf k}_0}(E) - I^-_{{\bf k}_0}(E)}{I^+_{{\bf k}_0}(E) + I^-_{{\bf k}_0}(E)}\,.
\end{equation}

In the calculations we again take a $50\times50\times50$  $k$-mesh in $1/8$ of the (cubic) Brillouin Zone. The results of the polarization \eqref{eq:P_E2} as a function of energy are shown in Figure~\ref{fig:PEs} by the dashed blue line. The mesh used for calculating the states and polarization is the same as for the effective phonon model.

\subsubsection{Analysis of the results}\label{sec:spinrel}

Let us discuss briefly the obtained results in the two simplified models plotted by the blue lines in Figure~\ref{fig:PEs}. Overall, the behavior of the optical orientation degree as function of detuning found from the model analysis is similar to that observed in experiment: up to a detuning of about 0.5~eV the optical orientation degree is practically constant and for larger detunings it drops. Both theoretical curves (at large detunings) are above the experimental data. This is because additional mechanisms of spin relaxation, namely, the Yafet contribution to the Elliott-Yafet mechanism~\cite{yafet63} and the Bir-Aronov-Pikus mechansism~\cite{BAP} are disregarded. They may contribute to the spin depolarization of electrons and holes. Let us analyze these effects in more detail.

To that end, we still consider a simplified approach, where the electrons and holes are excited, then loose their excess energy independently, and subsequently bind into excitons before recombination. We also assume that the electron-hole scattering is negligible and disregard the Bir-Aronov-Pikus mechanism of spin relaxation related to the electron-hole exchange,\cite{opt_or_book,BAP} see SI to Ref.~\citenum{kirstein2021} for a brief description of the Bir-Aronov-Pikus mechanism in perovskites. We assume that both for electrons and holes their momentum relaxation is fast compared to their spin and energy relaxation. As a result, the spin dynamics of each type of charge carrier can be described by a kinetic equation in the form 
\begin{equation}
\label{spin:rel}
\frac{d\mathbf s(\varepsilon)}{dt} + \frac{\mathbf s(\varepsilon)}{\tau_s(\varepsilon)} = Q_\varepsilon\{\mathbf s(\varepsilon)\}.
\end{equation}
Here $\varepsilon$ is the charge carrier energy reckoned from the conduction band bottom (for electrons) or valence band top (for holes), $\mathbf s(\varepsilon)$ is the energy-dependent spin distribution function of electrons or holes, $\tau_s(\varepsilon)$ is the energy-dependent spin relaxation time. $Q_\varepsilon\{\mathbf s(\varepsilon)\}$ is the phonon-carrier collision integral describing the energy relaxation. Within this simplified approach the additional effects of spin relaxation appear during thermalization of charge carriers~\cite{DP}:
\begin{equation}
\label{zeta}
\mathbf s(0) = \zeta(\Delta E) \mathbf s(\Delta E),
\end{equation}
where
\begin{equation}
\label{zeta:1}
 \zeta(\Delta E) = \exp{(-\Phi)}, \quad \Phi = \int_0^{\Delta E} \frac{d \varepsilon}{\varepsilon} \frac{\tau_\varepsilon(\varepsilon)}{\tau_s(\varepsilon)} \,.
\end{equation}  
Here $\tau_\varepsilon(\varepsilon)$ is the energy relaxation time. The depolarization factor $\zeta(\Delta E)$, where $\Delta E$ is the energy of the charge carrier above (below) the band bottom (top), is generally different for electrons and holes. 

The depolarization effect described by Eq.~\eqref{zeta} can be quite significant. Let us consider the generic and simplified case, when the detuning energy is much greater than the optical phonon energy, and take into account the scattering via this optical phonon only. In that case the energy relaxation occurs via emission of a cascade of optical phonons.  We take the Fr\"ohlich (piezo-optical) mechanism\cite{gantmakher87} of the electron-longitudinal optical phonon interaction for illustration (note that this mechanism is present in perovskites and is relatively strong~\cite{Wright:2016aa}). The energy relaxation rate due to the electron-phonon interaction with the phonon mode of energy $\hbar \Omega$ is given by 
\begin{equation}
\label{energ:PO}
\frac{1}{\tau_\varepsilon(\varepsilon)} = \frac{1}{\tau_{PO}} \left(\frac{\hbar\Omega}{\varepsilon}\right)^{3/2},
\end{equation} 
where $\tau_{PO} = (2\alpha \Omega)^{-1}$ is the characteristic electron (or hole) phonon interaction time with $\alpha$ being the Fr\"ohlich constant.\cite{gantmakher87} In the simplest model the spin relaxation rate within the Elliott-Yafet mechanism can be written as~\cite{opt_or_book}
\begin{equation}
\label{spin:DO}
\frac{1}{\tau_s(\varepsilon)} = \frac{C(\varepsilon)}{\tau_{PO}} \left(\frac{\hbar\Omega}{\varepsilon}\right)^{1/2}.
\end{equation}
Here the function $C(\varepsilon) = \xi (\varepsilon/E_g)^2$ provides the ratio of the spin-flip to the momentum relaxation rates owing to the generic $[\mathbf k \times \mathbf k']\mathbf s$ terms in the electron (hole)-phonon interaction Hamiltonian.  $\xi$ is a constant on the order of unity, which depends on the type of the charge carrier and is related to the strength of the spin-orbit coupling (in perovskites the spin-orbit splitting of the conduction band is on the order of the band gap~\cite{kirstein2021nc}). The exponent in Eq.~\eqref{zeta:1} then reads
\begin{equation}
\label{Phi:dep}
\Phi(\Delta E) = \frac{\xi}{3} \frac{(\Delta E)^3}{\hbar\Omega E_g^2}.
\end{equation}
This factor can be quite significant for $\Delta E = E- E_g \sim 1$~eV, resulting is a large depolarization, $\zeta(\Delta E) \to 0$. As a result, inclusion of the spin relaxation processes beyond the effective phonon model can account for the discrepancy between the simplified models above and the experiment.

To analyze the effect of spin relaxation quantitatively, we include the simplified model presented above in the calculations of the optical orientation by replacing $\rho^{\eta} ({\bf k}_R;{\bf k}_0)$ in Eqs.~(\ref{eq:Ipm_general},\ref{eq:I_E}) with the matrix ($n_{\eta}=\operatorname{Tr}\left\{\rho^{\eta} ({\bf k}_R;{\bf k}_0)\right\}$)
\begin{equation}\label{eq:rho_SR}
  \rho^{\eta,\mathrm{SR}} ({\bf k}_R;{\bf k}_0) = n_{\eta}+  
  \left[\rho^{\eta} ({\bf k}_R;{\bf k}_0) - n_{\eta} \right] 
  \exp\left(-\frac{\xi_{\eta}}3 \frac{|E_{\eta}({\bf k}_0)-E_{\eta}({\bf k}_R)| }{\hbar \Omega E_g^2} \right)\,.
\end{equation}
Then, Eqs.~(\ref{eq:Ipm_general},\ref{eq:I_E},\ref{eq:rho_SR}) give an estimate of the optical orientation degree $P^{\mathrm{SR}}(E)$. In Figure~\ref{fig:PEs} we show $P^{\mathrm{SR}}(E)$ by the green line, calculated with $\hbar\Omega=15$~meV and $\xi_{c}=\xi_{v}=1$.
The energy of the phonon mode is chosen based on typical energies of phonons responsible for maximum electron-phonon interaction (see e.g. Ref.~\citenum{Ponce2019}). The numerical results assume that $P^{\mathrm{SR}}(E)$ only weakly depends on the value of $\hbar\Omega$. While the polarization strongly depends on both values of $\xi$, we have found surprisingly good agreement between the experimental data and the calculated results for $\xi_{c}=\xi_{v}=1$ without any additional fitting.

Above, we discussed the depolarization of free carriers. The optical orientation degree of the emission is also influenced by the exciton formation processes. Here, two scenarios are possible [cf. Sec.~\ref{sec:correlations:end}]:
\begin{itemize}
\item Electrons and holes loose their energy independently and then excitons are formed from partially polarized charge carriers. In that case the optical orientation degree equals to the exciton polarization degree and takes the form~\cite{kocher98}
\begin{equation}
\label{exciton:pol}
P_X=\frac{P_e+P_h}{1+P_eP_h},
\end{equation}
where $P_e$ and $P_h$ are the polarization degrees of electrons and holes, respectively, calculated with account for the depolarization factors~\eqref{zeta} and \eqref{Phi:dep}. Note that in this way the polarization of excitons can be higher than the polarization of the individual carriers.
\item In the other scenario the partially polarized electrons and holes recombine with unpolarized long-living (resident) carriers. In that case the emission polarization will be given by the average of the electron and hole polarizations, provided a negligible resident carriers' polarization is present.
\end{itemize}

While the analysis above provides a qualitative and even semi-quantitative description of the optical orientation in perovskites for non-resonant excitation, we abstain from a detailed comparison between the models and the experiments. In fact, the derivation of the depolarization factor~\eqref{zeta} is correct only for not too high excess energies, where the Fr\"ohlich interaction can be described macroscopically and where the spin-orbit contributions to the scattering matrix element take a simple $[\mathbf k \times \mathbf k']\mathbf s$ form. Development of a full theory of spin and energy relaxation in perovskites goes beyond the present work.

\end{bibunit}
\end{document}